\documentclass[a4paper,onecolumn,unpublished,amsfonts,amsmath,amssymb,accepted=2024-06-14]{quantumarticle}
\pdfoutput=1
\usepackage[utf8]{inputenc}
\usepackage[english]{babel}
\usepackage[T1]{fontenc}
\usepackage[unicode=true,bookmarks=true,bookmarksopen=false,breaklinks=false,pdfborder={0 0 0},colorlinks=true]{hyperref}
\usepackage{xcolor}
\definecolor{cblue}{rgb}{0.16, 0.32, 0.75}
\definecolor{cred}{rgb}{0.7, 0.11, 0.11}
\hypersetup{%
	,linkcolor=cred
	,citecolor=cblue
	,urlcolor=cblue
}

\usepackage[numbers]{natbib}
\usepackage{dsfont,enumitem,color,bm,ulem}
\usepackage[makeroom]{cancel}
\usepackage[allowbreakbefore,nospacearound,shortcuts]{extdash}

\allowdisplaybreaks				
														
\begin{document}

\title{Objectivity of classical quantum stochastic processes}
\author{Piotr Sza\'{n}kowski}
\address{Institute of Physics, Polish Academy of Sciences, al.~Lotnik{\'o}w 32/46, PL 02-668 Warsaw, Poland}
\email{piotr.szankowski@ifpan.edu.pl}
\author{\L{}ukasz Cywi\'{n}ski}
\address{Institute of Physics, Polish Academy of Sciences, al.~Lotnik{\'o}w 32/46, PL 02-668 Warsaw, Poland}

\begin{abstract}
We investigate what can be concluded about a quantum system when sequential quantum measurements of its observable---a prominent example of the so-called quantum stochastic process---fulfill the Kolmogorov consistency condition and thus appear to an observer as a sampling of a classical trajectory. We identify a set of physical conditions imposed on the system dynamics, that when satisfied, lead to the aforementioned trajectory interpretation of the measurement results. We then show that when another quantum system is coupled to the observable, the operator representing it can be replaced by external noise. Crucially, the realizations of this surrogate (classical) stochastic process follow the same trajectories as those measured by the observer. Therefore, it can be said that the trajectory interpretation suggested by the Kolmogorov consistent measurements also applies in contexts other than sequential measurements. 
\end{abstract}

%\date{\today}
		
\maketitle

\section{Introduction}

Quantum mechanics is a flagship example of non-classical physical theory. To human users, who are firmly ingrained in the classical realm, many aspects of the theory often appear highly counter-intuitive and sometimes even paradoxical. This applies both to the elements of mathematical formalism and to the interpretative rules that correlate the abstract mathematics with user's experiences. As a result, achieving an intuitive understanding of a genuinely quantum phenomena poses significant challenge, especially when compared to classical theories.

One strategy for overcoming these difficulties is to identify scenarios where the quantum theory behaves classically and then explore what happens in the ``neighborhood''. To put it differently, the idea is to start at a more easily understandable, classical-like point and then gradually move towards a nearby non-classical aspect of the theory, one small step at a time. Additionally, reversing this approach---which, of course, requires locating the classical-like point in the first place---provides a method to investigate mechanisms governing the quantum--classical transition~\cite{Schlosshauer_book,Strasberg_SciPost23}. In this vein, recently, there has been resurgence of efforts in the search for the signatures of classicality manifesting in the statistics of the sequential quantum measurements \cite{Smirne_QST19,Strasberg_PRA19,Milz_PRX20,Szankowski_PRA21,Sakuldee_PRA21,Lonigro_Frontiers2022}.

In the language of quantum theory, the result of an experiment consisting of $n$ consecutive measurement events is depicted as a sequence of stochastic variables described by a joint probability distribution $P_n$---the \textit{quantum stochastic process}~\cite{Davies_CommMathPhys69}. Classical intuition suggests that sequential measurements should allow the observer to uncover a trajectory that was traced by the evolving physical quantity (the observable); after all, this is how measurements are supposed to work in classical physics. However, quantum observables often behave in ways that contradict this naive intuition.

Probability theory asserts that the measured sequence can be interpreted as a sampling of an underlying trajectory---i.e., a trajectory that was traced over time independently of the measurement events---only if the joint probability distributions satisfy the \textit{Kolmogorov consistency} (KC) condition. Formally, let $\Omega(F)$ be the set of all possible results of a single measurement of a quantum observable $F$. Then, $P_n(f_n,t_n;\ldots;f_1,t_1)$ is the probability of obtaining the sequence of results $f_n,\ldots,f_1$ [where each $f_i\in\Omega(F)$] in consecutive measurements performed at the corresponding times $0<t_1<\cdots<t_n$. The family $\{P_n\}_{n=1}^\infty$ satisfies KC condition (or ``is consistent'', for short) when
\begin{align}\label{eq:Born_KC}
    &\begin{aligned}[c]\text{\textbf{KC:}}&\end{aligned}
    \begin{aligned}[c]
          &\forall(n>1)\forall(0<t_1<\cdots<t_n)\forall(1\leq i \leq n)\forall(f_1,\ldots,f_{i-1},f_{i+1},\ldots,f_n \in\Omega(F) ) : \\
          &\!\!\!\!\sum_{f_i\in\Omega(F)}\!\!
            P_n(f_n,t_n; \ldots ; f_i, t_i; \ldots; f_1,t_1)=P_{n-1} (f_n,t_n; \ldots ; f_{i+1}, t_{i+1};f_{i-1},t_{i-1}; \ldots ; f_1,t_1).
    \end{aligned}
\end{align}
By the Kolmogorov extension theorem \cite{Kolmogorov}, any family of consistent joint probability distributions uniquely defines a stochastic process. This makes the sequence of measured results effectively indistinguishable from the sampling of a randomly chosen trajectory (the realization of the stochastic process), and thus, the quantum stochastic process is transmuted into the \textit{classical quantum process}. However, this is typically not the case because probability distributions $P_n$ describing quantum measurements generally violate KC condition \cite{Milz_Quantum20,Milz_PRXQ21} and fulfill it only under specific conditions \cite{Smirne_QST19,Strasberg_PRA19,Milz_PRX20,Szankowski_PRA21,Sakuldee_PRA21,Lonigro_Frontiers2022}.

One widely discussed example \cite{Smirne_QST19,Strasberg_PRA19,Milz_PRX20} is that of a Markovian open system, where it is assumed that $P_n$ can be parameterized with \textit{quantum regression formula}~\cite{Lonigro_JPA22,Lonigro_PRA22}. In this case, the KC condition translates into the constraint placed onto the assumed CP-divisible map representing the dynamical law, which could, in principle, be satisfied by some physical system. Another example was identified in a chaotic system~\cite{Strasberg_PRA23,Strasberg_SciPost23,Strasberg_arXiv23_2} where it was shown that measurements of sufficiently ``slow and coarse-grained'' observables satisfy the KC condition.

Here, we restate the problem of consistency using the formalism that allows for a unified treatment of these examples, as well as introduction of new ones. Overall, the rule of thumb is that KC is satisfied for measurements of observables in systems that are macroscopic, chaotic, subjected to the specific kind of decoherence, or a combination of these factors. Conversly, except for a handful of trivial examples, observables of microscopic, largely isolated, and integrable quantum system will almost always violate KC.

Therefore, in the special event when $\{P_n\}_{n=1}^\infty$ is consistent, the sequence of results witnessed by the classical observer can be modeled with a stochastic process. This means an observer performing the sequential quantum measurement of the observable represented by Hermitian operator $\hat F$ cannot distinguish their results from the sequential sampling of trajectory $f(t)$ that realizes the stochastic process $F(t)$.

Given this, one is compelled to pose the following question: can this trajectory picture be applied in contexts other than measurements performed by a classical observer? Can these trajectories be considered \textit{objective} entities? Answering such questions is challenging. In quantum mechanics, the classical observer occupies a special role, distinct from the rest of the quantum world, because in the standard formulation the measurement process is exempted from the ordinary dynamical laws of quantum systems. Consequently, it is generally assumed that the observer's perception of the measured system is a product of the measurement itself, and thus is inherently subjective.

To consider whether a given picture seen by the classical observer could be more than just subjective, it is necessary to compare their perceptions with the perceptions of \textit{non-classical observes} residing outside the measurement context. This poses significant conceptual challenge: to make such a comparison meaningful, one must first quantify the perceptions of non-classical observer and then find a way to place the measurement and non-measurement contexts on equal footing.

In the case study presented here, these abstract considerations are made more concrete, making the problem itself more approachable. The question of objectivity we pursue can be phrased as follows: if the classical observer perceives the observable $F$ as a stochastic process, can we say that also a non-classical observer---i.e., a quantum system coupled to this observable---also evolves as if the operator representing the observable were replaced with an external field represented by the process $F(t)$? If answer is affirmative, then we can say that the quantum measurement of the observable $F$ achieves the classical ideal where it becomes possible to observe the ``actual, objective state (the trajectory) of a physical quantity''. This is because, if there is a symmetry between classical and non-classical perceptions, the same fundamental description of the measurement result (the family $\{P_n\}_{n=1}^\infty$) also describes the evolution of other quantum systems coupled to $F$.

Here, we define the conditions the dynamics of the observable has to satisfy to achieve this objectivity of its trajectory picture and we discuss the physical implications. These conditions turn out to be formally equivalent to the decoherent histories criterion of the consistent histories framework~\cite{Griffiths_JSP84,Gell-Mann_PRD93,Strasberg_SciPost23,Strasberg_arXiv23_2}. However, the fact that they guarantee the above-defined objectivity of the trajectory picture has apparently gone unnoticed until now.

\section{Classical observer}\label{sec:classical_pov}

The classical observer experiences the dynamics of a quantum system through readouts of the dedicated \textit{measuring apparatus}. According to the standard interpretation of quantum mechanics, the apparatus itself is treated as a primitive notion: it is assumed that the device can be built, but the details of its inner workings and the particularities of its interactions with both the observer and the measured system are not specified. The one essential aspect that must be specified for every apparatus is the assignment of the Hilbert space partitioning---also known as a positive operator-valued measure---$\{\hat E(n)\}_n$, such that $\hat E(n)\geqslant 0$ and $\sum_n \hat E(n) = \hat 1$.

Partitionings into mutually orthogonal subspaces are of particular interest because they correspond to direct measurements of physical quantities (the observables). For example, the observable $F$ is represented in the formalism by the Hermitian operator $\hat F = \hat F^\dagger$ (the circumflex $(\,\hat{}\,)$ indicates the operator representation); this operator has a spectral decomposition,
    \begin{align}
        \hat F = \sum_{f\in\Omega(F)}f \hat P(f),
    \end{align}
where the set $\Omega(F)$ contains all unique eigenvalues of $\hat F$, and the operators $\hat P(f)$ are the projectors onto the corresponding orthogonal eigenspaces, i.e., $\sum_f\hat P(f) = \hat 1$, $\hat P(f)\hat P(f') = \delta_{f,f'}\hat P(f)$, and thus $\hat F\hat P(f) = f\hat P(f)$. (We will suppress the range of sums whenever it is clear from context that the variable belongs to $\Omega(F)$.) Therefore, $\{\hat P(f)\}_{f\in\Omega(F)}$ defines an orthogonal partitioning of the Hilbert space, which can then be assigned to the apparatus measuring the physical quantity $F$.

The \textit{Born rule} defines how the partitionings assigned to measuring apparatuses correlate with the readouts perceived by the classical observer. It states the following: given the unitary evolution operator $\hat U(t) = {\exp}(-it\hat H)$ describing the dynamical law in the measured system $S$ and the initial density matrix $\hat\rho$, the probability of reading out the result $f_1$ at the time $t_1 > 0$, and the result $f_2$ at $t_2>t_1$, \ldots, and finally the result $f_n$ at $t_n>t_{n-1}$, in the sequence of $n$ measurements is calculated according to the formula
\begin{align}
\label{eq:Born_rule}
    P_n(\mathbf{f}_n,\mathbf{t}_n)& = P_n(f_n,t_n;\ldots;f_1,t_1) 
    = \operatorname{tr}\left[\left(\prod_{i=n}^1\hat P(f_i,t_i)\right)\hat\rho\left(\prod_{i=1}^n \hat P(f_i,t_i)\right)\right].
\end{align}
Here, the symbol $\prod_{i=n}^1\hat A_i$ ($\prod_{i=1}^n\hat A_i$) indicates an ordered composition $\hat A_n\allowbreak \cdots \allowbreak\hat A_1$ ($\hat A_1\cdots\hat A_n$), and $\hat P(f,t) = \hat U^\dagger(t)\hat P(f)\hat U(t)$ is the Heisenberg picture of the projector. We also adopt a shorthand notation where a quasi-vectors of the form $\mathbf{k}_n = (k_n,\ldots,k_1)$, denoted by the lowercase boldface and the subscript indicating the number of elements, are used as a stand-in for the sequence of arguments ranging from $k_n$ to $k_1$.

By default, the \textit{Born distributions} $P_n$ are \textit{causal}~\cite{Milz_Quantum20},
\begin{align}\label{eq:collapse_rule}
    \forall (n \geqslant 1)\forall\mathbf{f}_n\,\forall(0<t_1<\cdots<t_n < t):
    \sum_{f} P_{n+1}(f,t;f_n,t_n;\ldots;f_1,t_1) = P_{n}(\mathbf{f}_n,\mathbf{t}_n).
\end{align}
Note the difference between the causality and the Kolmogorovian consistency (KC): KC requires that marginalization over any argument reduces the order $n$, while in the case of Eq.~\eqref{eq:collapse_rule} the reduction is guaranteed only when marginalizing over the latest argument. Overall, the causality is not related to consistency---it is a unique property of Born distributions that allows to rewrite joint distributions $P_n$ as a product of conditional probabilities,
\begin{align}\label{eq:Gleason_dist}
    P(f|\hat\rho_t) = \operatorname{tr}[\hat P(f)\hat\rho_t],
\end{align}
with the condition in a form of density matrix $\hat\rho_t$,
\begin{align}
\nonumber
    P_n(\mathbf{f}_n,\mathbf{t}_n) &= \frac{P_n(\mathbf{f}_n,\mathbf{t}_n)}{P_{n-1}(\mathbf{f}_{n-1},\mathbf{t}_{n-1})}
        \frac{P_{n-1}(\mathbf{f}_{n-1},\mathbf{t}_{n-1})}{P_{n-2}(\mathbf{f}_{n-2},\mathbf{t}_{n-2})}\cdots\frac{P_2(\mathbf{f}_2,\mathbf{t}_2)}{P_1(f_1,t_1)}P_1(f_1,t_1)\\
\label{eq:collapse_form}
    &= P(f_1|\hat U(t_1)\hat\rho\hat U^\dagger(t_1))\prod_{k=1}^{n-1} P(f_{k+1}|\hat\rho_{t_{k+1}|\mathbf{f}_{k},\mathbf{t}_{k}}).
\end{align}
where the density matrices conditioned on the history of previous measurement results are given by
\begin{align}
    \hat\rho_{t_{k+1}|\mathbf{f}_k,\mathbf{t}_k} &= \hat U(t_{k+1})\frac{\left(\prod_{i=k}^1\hat P(f_i,t_i)\right)\hat\rho\left(\prod_{i=1}^k \hat P(f_i,t_i)\right)}{P_k(\mathbf{f}_k,\mathbf{t}_k)}\hat U^\dagger(t_{k+1}),
\end{align}
and the relation~\eqref{eq:collapse_rule} ensures their proper normalization. Typically, distribution of form~\eqref{eq:Gleason_dist} is interpreted as describing a single measurement performed on the current \textit{state of the system} $\hat\rho_t$~\cite{Gleason_57}; therefore, the reformulation~\eqref{eq:collapse_form} suggests a reading that the sequential measurement is a composition of independent measurement events, where in each step the apparatus measures the state as it is at the corresponding instant. However, it looks as if the very act of observation changes the state in between the steps. Indeed, we see in Eq.~\eqref{eq:collapse_form} that the state at the time of the next event, $\hat\rho_{t_{k+1}|\mathbf{f}_k,\mathbf{t}_k}$, appears as the state from the previous event, $\hat\rho_{t_k|\mathbf{f}_{k-1},\mathbf{t}_{k-1}}$, which was \textit{collapsed} onto the subspace corresponding to the measured result, 
\begin{align}
    \hat\rho_{t_{k+1}|\mathbf{f}_k,\mathbf{t}_k} &= \hat U(t_{k+1}-t_k)\frac{\hat P(f_k)\hat \rho_{t_k|\mathbf{f}_{k-1},\mathbf{t}_{k-1}}\hat P(f_k)}{P(f_k|\hat \rho_{t_k|\mathbf{f}_{k-1},\mathbf{t}_{k-1}})}\hat U^\dagger(t_{k+1}-t_k).
\end{align}
Thus, in this reading, the statistical dependence between results in a sequential measurement is shifted to the state subjected to collapse events coinciding with each observation.

Can this picture of the system state being collapsed by the measurement be objectivized? What standard should one apply to decide whether Eq.~\eqref{eq:collapse_form} and the accompanying narrative are sufficient reasons to consider state collapse as an objective event? Notice that the very idea of collapse, understood as a sudden, non-unitary (i.e., not following the dynamical law of the system) change of state that coincide with the measurement event, arose solely because of specific algebraic transformations to the formula defining $P_n$. Given that, the most that can be said at this point is that the collapse is something subjectively perceived by a classical observer performing sequential measurements. To even entertain the possibility of its objectivization, it should be possible to show this kind of state change appearing in at least one context other than the sequential measurements. The more diverse the contexts the picture of collapsing states can be adopted, the more justified the claim of its objectivity~\cite{Nagel_book}---this is the standard we are setting for our present investigations. To our best knowledge, collapse has not been demonstrated in any other context of the quantum theory, and so, the state collapse picture does not meet our standard. Therefore, the collapse is, at most, \textit{inter-subjective} among classical observers. The term ``inter-subjectivity'' is used here in the sense that the subjective perceptions of classical observers measuring the observable $F$ are in agreement with each other. We can say that those observers are indeed in agreement because the same family of Born distributions is used to describe perceptions of each one of them. Our aim is to show that the trajectory picture emerging when Born distributions are consistent turns out to be more than inter-subjective when assessed against our standard of objectivity---we will demonstrate that the trajectory picture can be applied in contexts beyond sequential measurements.

\section{Consistent Measurements}\label{sec:consistent_measurmements}

The causality relation~\eqref{eq:collapse_rule} applies only to the latest measurement result in the sequence; analogous relations between Born distributions involving mid-sequence results are not readily apparent in the general case. Nevertheless, quantum mechanics does not prohibit such relations as a matter of principle. An important example---and the subject of this paper---is the Kolmogorov consistency~\eqref{eq:Born_KC} (KC), a relation between Born distributions inspired by the fundamental laws of the classical theory of stochastic processes~\cite{Kolmogorov,vanKampen_book}.
    
In classical theory, physical quantities are represented by trajectories traced over time in accordance with the system's dynamical laws. These trajectories are considered objective entities in the sense that the same fundamental description of the trajectory accounts for both measurement results and any other non-measurement interaction with the physical quantity in question: ``when the (classical) tree falls, it makes a sound even if there is no one around to hear it.'' The Kolmogorovian consistency is a manifestation of this foundational postulate of classical physics. Let us explain why by reviewing some fundamentals of the theory of stochastic processes. 

Formally, a classical stochastic process $X_\mathrm{cls}(t)$ representing a physical quantity is defined as a map from the set of outcomes of random event (the sample space) to \textit{trajectories} $x(t)$---the real-valued functions of time $t\in\mathbb{R}$. Every process $X_\mathrm{cls}(t)$ is assigned a probability distribution functional $P^{X_\mathrm{cls}}[x]$ for its trajectories $x(t)$; as a probability distribution, this functional is non-negative, $P^{X_\mathrm{cls}}[x]\geqslant 0$, and normalized,
    \begin{align}
        \int P^{X_\mathrm{cls}}[x][Dx]=1,
    \end{align}
where $\int\cdots[Dx]$ indicates the functional integration. Given the probability distribution $P^{X_\mathrm{cls}}[x]$, one can calculate the expectation value of arbitrary functionals of $X_{\mathrm{cls}}(t)$,
    \begin{align}\label{eq:stochastic_avg}
        \overline{W[X_\mathrm{cls}]} = \lim_{N\to\infty}\frac{1}{N}\sum_{i=1}^N W[x_i] = \int P^{X_\mathrm{cls}}[x]W[x][Dx],
    \end{align}
where $\{x_i(t)\}_{i=1}^N$ is an ensemble of independently sampled trajectories. The \textit{moments} of the process,
    \begin{align}\label{eq:moment_trajectory_avg}
        \overline{X_\mathrm{cls}(t_n)\cdots X_\mathrm{cls}(t_1)} &= \int P^{X_\mathrm{cls}}[x]\left(\prod_{i=1}^n x(t_i)\right)[Dx],
    \end{align}
are important examples of expectation values. Since any regular functional $W[X_\mathrm{cls}]$ has some form of series expansion into a combination of products of $X_\mathrm{cls}(t)$, the problem of computing $\overline{W[X_\mathrm{cls}]}$ can be broken down into manageable steps each solved with the use of an appropriate moment. However, the functional integral form of stochastic average~\eqref{eq:moment_trajectory_avg} is difficult to work with. In practical applications, the formal definition utilizing the functional $P^{X_\mathrm{cls}}[x]$ is almost always rewritten in the language of \textit{joint probability distributions} $\{P_n^{X_\mathrm{cls}}\}_{n=1}^\infty$. Each joint distribution is a standard function of $n$ pairs of arguments: a real value $x_i$ and the time $t_i$. The functions $P_n^{X_\mathrm{cls}}$ are proper multivaried probability distributions that are non-negative and normalized,
    \begin{align}
    \nonumber
        &\forall(n\geqslant 1)\forall(x_1,\ldots,x_n)\forall(0<t_1<\cdots<t_n):\\
        &\phantom{=}
        P^{X_\mathrm{cls}}_n(x_n,t_n;\ldots;x_1,t_1) \geqslant 0;\quad
        \sum_{x_n,\ldots,x_1}P_n^{X_\mathrm{cls}}(x_n,t_n;\ldots;x_1,t_1) = 1.
    \end{align}
(For simplicity, we are assuming here that process $X_\mathrm{cls}(t)$ is discrete, hence the sums over $x_i$ rather than integrals, although $t_i\in\mathbb{R}$.) The joint distribution $P_n^{X_\mathrm{cls}}(x_n,t_n;\ldots;x_1,t_1)$ is interpreted as the probability that a randomly chosen trajectory $x(t)$ passed through all the consecutive values $x_i$ at the corresponding points in time $t_i$ (assuming $t_1<\cdots<t_n$); in formal terms,
    \begin{align}\label{eq:process_joint_dist}
        P_n^{X_\mathrm{cls}}(x_n,t_n;\ldots;x_1,t_1) &= \int P^{X_\mathrm{cls}}[x]\left(\prod_{i=1}^n\delta(x(t_i)-x_i)\right)[Dx].
    \end{align}
This means that moments (and thus, any expectation value) can be rewritten to utilize the joint distributions instead of the unwieldy functional integration,
    \begin{align}
        \int P^{X_\mathrm{cls}}[x]\left(\prod_{i=1}^n x(t_i)\right)[Dx]
        &= \sum_{x_n,\ldots, x_1}P_n^{X_\mathrm{cls}}(x_n,t_n;\ldots;x_1,t_1)\left(\prod_{i=1}^n x_i\right).
    \end{align}
However, as the joint probability distributions are meant to describe the course of objective trajectories, they have to be \textit{consistent}. If $P_n^{X_\mathrm{cls}}(\ldots;x_i,t_i;\ldots)$ is the probability that trajectory has passed through $x_i$ at time $t_i$, then $P_n^{X_\mathrm{cls}}(\ldots;x_i,t_i;\ldots)+P_n^{X_\mathrm{cls}}(\ldots,x_i',t_i;\ldots)$ is the probability that $x(t)$ has passed through either $x_i$ or $x_i'$, and thus, $\sum_{x_i}P_n^{X_\mathrm{cls}}(\ldots;x_i,t_i;\ldots)$ is the probability that it has passed through any value. Since trajectories are functions defined on the whole real line, the probability for the occurrence of all possible alternatives at time $t_i$ (i.e., trajectory passing through any value) must be equal to the probability without any constraints imposed on the trajectory at this point in time. But the probability for the trajectory to pass through the sequence of values $x_n,\ldots,x_1$ minus the constraint at $t_i$ is given by the joint distribution of order $n-1$ where the $i$th pair of arguments is skipped, $P_{n-1}^{X_\mathrm{cls}}(\ldots;x_{i+1},t_{i+1};x_{i-1},t_{i-1};\ldots)$---i.e., the Kolmogorov consistency. Hence, in the context of classical theory, the fact that stochastic processes are realized as trajectories is manifested as consistency of joint probability distributions. The reciprocal is also true: the extension theorem states that any infinite family of multivaried probability distributions that is consistent, uniquely defines a stochastic process with trajectories as its realizations. In other words, if a given family of joint probability distributions satisfy the KC condition, then the extension theorem asserts that each member of the family is a restriction [in the sense of Eq.~\eqref{eq:process_joint_dist}] of a ``master'' probability distribution functional.

Coming back to the issue of quantum observables, when the system's properties cause the Born distributions to satisfy KC condition, the extension theorem implies that $\{P_n\}_{n=1}^\infty$ defines stochastic process $F(t)$. Distributions $P_n$ then play the role of joint probabilities $P_n^F$, and, as for all stochastic processes, they combine into functional distribution $P^F[f]$ for process trajectories $f(t)$. Consequently, each value in a sequence $f_n,\ldots,f_1$ [with probability distribution $P_n(f_n,t_n;\ldots;f_1,t_1)$] is equivalent to the sample of a trajectory, $f_i = f(t_i)$, where $f(t)$ has distribution $P^F[f]$. On a surface level, we conclude from these observations that the results of the sequential measurements of KC-satisfying quantum observable $F$ can be simulated with the sampling of a classical stochastic process $F(t)$ without the loss of any information. On a higher level, the KC condition~\eqref{eq:Born_KC} itself suggests the interpretation that $F$ appears to classical observer as a trajectory traced over time independently of the action of measuring apparatus, and the individual measurement events only uncover the already determined value.

To explain how we are coming to this conclusion, first consider the following. In the theory of probability, the sum over $i$th argument on the LHS of KC condition~\eqref{eq:Born_KC} is typically meant to indicate that the measurement was performed at time $t_i$ but the observer has discarded or forgotten the result. On the other hand, the Born distribution that skips the argument at $t_i$ describes the situation where there was no measurement at that time. Hence, the KC condition could be understood to mean that the influence of the measuring apparatus on the course of the observable dynamics is insignificant (in the sense that it does not affect the statistics of the following measurements), because simply forgetting the result of a measurement that was performed is indistinguishable from the situation when the measurement never happened in the first place. 

In summary, we have argued here that in the special event when the Born distributions $\{P_n\}_{n=1}^\infty$ satisfy the Kolmogorov consistency condition, the classical observers perceive the measured quantum observable as a trajectory akin to realizations of stochastic process. In such a case, we say that the \textit{trajectory picture} (of observable $F$) applies in the context of sequential measurements. Therefore, when tested against our standard of objectivity defined in the previous Sec.~\ref{sec:classical_pov}, at this point, we can only affirm that the trajectory picture is, at least, inter-subjective among classical observers.

However, in what follows we will show that the trajectory picture is more than inter-subjective because---unlike the previously discussed state collapse picture---it \text{can} be applied in context that do not involve measurements or classical observers. To this end, we shall first define the notion of \textit{non-classical observer} and introduce the formalism that allows to speak of its perceptions of quantum observables.

\section{Non-classical observer}\label{sec:non-classical_pov}

We define the non-classical observer (of the observable $F$) as any other quantum system $O$, with its own dynamical law $\hat U_o(t) = {\exp}(-i t\hat H_o)$, that is brought into contact with the original system $S$. For $O$ to ``observe'' the physical quantity $F$ (analogous to classical observer performing measurements of $F$ directly on $S$), the two systems need to interact according to the law involving the operator representation $\hat F$, i.e., the total $OS$ Hamiltonian of the form
\begin{align}\label{eq:H_OS}
    \hat H_{os}= \hat H_{o}\otimes\hat 1 + \hat 1\otimes\hat H + \lambda \hat G_o\otimes\hat F,
\end{align}
with the coupling strength $\lambda$ and arbitrary $\hat H_o$, $\hat G_o$, is a minimal model of such an ``observing'' interaction. The unitary evolution generated by this minimal model reads
\begin{align}\label{eq:U_os}
    \hat U_{os}(t) &= e^{-it\hat H_{os}}= \hat U_o(t)\otimes\hat U(t) \hat V_{os}(t,0),
\end{align}
where the interaction picture evolution operator is given by the standard time-ordered exponential,
\begin{align}
\label{eq:SO_int}
    \hat V_{os}(t,s) &= Te^{-i\lambda\int_s^t\hat G_o(\tau)\otimes\hat F(\tau)d\tau}
    = \sum_{k=0}^\infty (-i\lambda)^k \!\!\int\limits_s^t\!\!d\tau_k\cdots\!\!\int\limits_s^{\tau_2}\!\!d\tau_1
    \prod_{i=k}^1\hat G_o(\tau_i)\otimes\hat F(\tau_i),
\end{align}
and the interaction pictures of coupling operators are given by $\hat G_o(\tau) = \hat U_o^\dagger(\tau)\hat G_o\hat U_o(\tau)$ and $\hat F(\tau) = \hat U^\dagger(\tau)\hat F\hat U(\tau)$. The system $O$ is, of course, changed by the interaction with $\hat F$, which manifests as deviations from the system's free evolution. The most basic way to quantify these changes, and thereby lay the foundations for the concept of non-classical perceptions, is to examine the dynamics of the reduced state of $O$, $\hat\rho_o(t) = \operatorname{tr}_s[\hat U_{os}(t)\hat\rho_o\otimes\hat\rho\,\hat U_{os}^\dagger(t)]$, or more precisely, its interaction picture $\hat\varrho_o(t) = \hat U_o^\dagger(t)\hat\rho_o(t)\hat U_o(t)$. 

To discuss the perceptions of $F$ for non-classical observers in concrete terms, we need to adopt an adequate formalism. For this purpose, we will utilize the so-called \textit{bi-probability parameterization} introduced previously as a tool for characterizing noise representations of open system dynamics~\cite{Szankowski_SciRep20,Szankowski_SciPostLecNotes23_}. The parameterization consists of the family of \textit{bi-probabilities} $\{Q_n^F\}_{n=1}^\infty$ associated with a given observable, $F$ in our case, with members defined as
\begin{align}\label{eq:Q}
    Q_n^F(\mathbf{f}_{n},\mathbf{f}_{-n},\mathbf{t}_n) &= Q_n^F(f_{n},f_{-n},t_n;\ldots;f_{1},f_{-1},t_1)= \operatorname{tr}\left[
        \left(\prod_{i=n}^1\hat P(f_{i},t_i)\right)\hat\rho\left(\prod_{i=1}^n \hat P(f_{-i},t_i)\right)
    \right],
\end{align}
where we use the same notation convention as in Sec.~\ref{sec:classical_pov}. Each bi-probability $Q_n^F$ is a complex-valued function of $n$ argument triplets: the time point $t_i$, and a pair of real eigenvalues $f_{\pm i}$. Using the decomposition of identity, $\sum_f \hat P(f,t) = \hat 1$, it is straightforward to verify through direct calculation that bi-probabilities satisfy a relation analogous to classical Kolmogorovian consistency,
\begin{align}\label{eq:quasi-KC}
\nonumber
    &\forall(n>1)\forall(0<t_1<\cdots<t_n)\forall(1\leq i \leq n)
        \forall(f_{\pm 1},f_{\pm(i-1)},f_{\pm(i+1)},f_{\pm n}) : \\
\nonumber
    &\sum_{f_{i},f_{-i}} Q_n^F(\mathbf{f}_{n},\mathbf{f}_{-n},\mathbf{t}_n)\\
    &\phantom{===}   
        = Q_{n-1}^F(f_{n},f_{-n},t_n;\ldots;f_{i+1},f_{-(i+1)},t_{i+1};f_{i-1},f_{-(i-1)},t_{i-1};\ldots;f_{1},f_{-1},t_1).
\end{align}
When compared with the role consistency plays in the classical theory of stochastic processes, this \textit{bi-consistency} suggests interpreting sequences $\mathbf{f}_{\pm n} = (f_{\pm n},\ldots,f_{\pm 1})$ as discrete samples of an underlying pair of trajectories $(f_+(t),f_-(t))$. Therefore, it is permissible to symbolically combine the family of bi-probabilities $\{Q_n^F\}_{n=1}^\infty$ into a single master functional distribution $Q^F[f_+,f_-]$, in the sense that $Q_n^F$ is considered the $n$-point restriction of $Q^F$,
\begin{align}\label{eq:gen_extension_thrm}
    Q_n^F(\mathbf{f}_n,\mathbf{f}_{-n},\mathbf{t}_n) &= \iint Q^F[f_+,f_-]\Big(\prod_{i=1}^n\delta(f_+(t_i)-f_i)\delta(f_-(t_i)-f_{-i})\Big)[Df_+][Df_-].
\end{align}
Doing so enables a convenient notation scheme that formally resembles the stochastic average formula~\eqref{eq:stochastic_avg} where, analogous to stochastic process $X_\mathrm{cls}(t)$ realized by the trajectory $x(t)$, the trajectory pair $(f_+,(t),f_-(t))$ would be a realization of a two-component \textit{bi-stochastic process} $(F_+(t),F_-(t))$,
\begin{align}
    \big\langle W[F_+,F_-]\big\rangle &=\iint Q^F[f_+,f_-]W[f_+,f_-][Df_+][Df_-],
\end{align}
so, for example, we can write
\begin{align}
    Q_n^F(\mathbf{f}_n,\mathbf{f}_{-n},\mathbf{t}_n) = \Big\langle \prod_{i=1}^n\delta(F_+(t_i)-f_i)\delta(F_-(t_i)-f_{-i})\Big\rangle.
\end{align}
We will make extensive use of this notation convention.\footnote{
While this manuscript was undergoing the review process, the proof of the generalized Kolmogorov extension theorem for bi-probabilities was reported in~\cite{Lonigro_24}. Consequently, it can now be said that the master functional $Q^F[f_+,f_-]$ does indeed exist, and the relation~\eqref{eq:gen_extension_thrm} is a mathematical fact rather than just a notation convention.
} To conclude this brief introduction to bi-probability formalism, we note that Eq.~\eqref{eq:Q} resembles the formula for the so-called \textit{decoherence functional} found in the consistent histories formulation of quantum mechanics~\cite{Griffiths_JSP84,Gell-Mann_PRD93,Strasberg_SciPost23,Strasberg_arXiv23_2}; we defer a more detailed discussion of this link to the end of this section.

The first major payoff from employing this formalism is the ability to capture, with bi\-/probabilities, the totality of the influence exerted by system $S$ onto the observer system $O$. It was shown in~\cite{Szankowski_SciRep20,Szankowski_SciPostLecNotes23_} that the reduced density matrix $\hat\varrho_o(t)$ can be rewritten in terms of \textit{bi-average} $\langle\ldots\rangle$, where the instances of observable operator $\hat F(t)$ (and any other influences from system $S$) are represented by the components of bi-stochastic process $(F_+(t),F_-(t))$, averaged over its bi-probability distributions $\{Q_n^F\}_{n=1}^\infty$:
\begin{align}
\label{eq:observer_state}
    \hat\varrho_o(t) &= \operatorname{tr}_s\left[\hat V_{os}(t,0)\hat\rho_o\otimes\hat\rho\,\hat V_{os}(0,t)\right]
        =\left\langle \hat V_o[F_{+}](t,0)\hat\rho_o\hat V_o[F_{-}](0,t)\right\rangle,
\end{align}
where the auxiliary $O$-only operator functional is derived from the $OS$ interaction operator, Eq.~\eqref{eq:SO_int}, by replacing $\hat F(\tau)$ with a function,
\begin{align}
    \hat V_o[\varphi](t,s) &\equiv Te^{-i\lambda\int_s^t \varphi(\tau)\hat G_o(\tau)d\tau}
     = \sum_{k=0}^\infty (-i\lambda)^k\!\!\int\limits_s^t\!\!d\tau_k\cdots\!\!\int\limits_s^{\tau_2}\!\!d\tau_1 \prod_{i=k}^1 \varphi(\tau_i)\hat G_o(\tau_i).
\end{align}

For completeness, the full step-by-step derivation of this result is presented in Appendix~\ref{appx:quasi-average_form}.

The second advantage of using the bi-probabilities is their meaningful comparison with the formalism of Born probabilities that describe the perceptions of observable $F$ for classical observers. To see this, it is crucial to realize that the family of bi-probabilities can be considered as a standalone object, abstracted from the specific context of dynamics of the reduced state of $O$. In such an approach, bi-probabilities quantify the dynamics of a given observable $F$. Using formula~\eqref{eq:Q}, one can compute the bi-probability $Q_n^F$ associated with the chosen observable given its spectral decomposition $\{\hat P(f)\}_f$, the dynamical law $\hat U(t)$ for the system the observable lives in, and the initial condition $\hat\rho$. Appendix~\ref{sec:SFR_examples} showcases several examples where $\{Q_n^F\}_{n=1}^\infty$ have been computed from first principles for a collection of concrete quantum systems. 

As a standalone object, the family of bi-probabilities contains the description of the classical observer's perceptions. Indeed, note that the \textit{diagonal} part of $Q_n^F$ (when both sequences of arguments overlap, $\mathbf{f}_{-n}=\mathbf{f}_n$) equals the Born distribution for the sequential measurement of observable $F$:
\begin{align}\label{eq:Q_diag}
    Q_n^F(\mathbf{f}_n,\mathbf{f}_n,\mathbf{t}_n) &= P_n(\mathbf{f}_n,\mathbf{t}_n)    
\end{align}
However, the ``classical'' Born distribution is only a part of the whole bi-probability. Since the two components of bi-stochastic process both appear in~\eqref{eq:observer_state}, the bi-average form of the reduced state demonstrates that both the diagonal and off-diagonal parts of bi-probabilities come into play in the context of the state dynamics of the quantum observer $O$. Moreover, this result generalizes to any context involving physical quantities (observables) in $O$; the influence from $F$ onto bi-probabilities associated with an arbitrary $O$-only observable is fully described by $\{Q_n^F\}_{n=1}^\infty$ as well. For example, the (stand-alone) bi-probabilities associated with observable represented by $\hat X_o = \sum_x x\hat P_o(x)$ can also be expressed in terms of a bi-average (see appendix~\ref{appx:quasi-average_form} for detailed derivation):
\begin{align}
\nonumber
    &Q_n^{X_o}(\mathbf{x}_n,\mathbf{x}_{-n},\mathbf{t}_n)\\
\label{eq:Q_quasi-avg_form}
    &\phantom{=}
        =\operatorname{tr}\left\langle
            \left(\prod_{i=n}^1\hat V_o[F_+](0,t_i)\hat P_o(x_i,t_i)\hat V_o[F_+](t_i,0)\right)\hat\rho_o
            \left(\prod_{i=1}^n\hat V_o[F_-](0,t_i)\hat P_o(x_{-i},t_i)\hat V_o[F_-](t_i,0)\right)
        \right\rangle.
\end{align}

Based on these observations, we conclude that just as Born distributions $\{P_n\}_{n=1}^\infty$ describe how classical observers perceive the observable $F$, bi-probabilities $\{Q_n^F\}_{n=1}^\infty$ describe perceptions of $F$ for non-classical observers (i.e.,~other quantum systems coupled to operator $\hat{F}$). The plural ``observers'' here is intentional because $Q_n^F$'s are determined exclusively by the properties of the observed system $S$, hence one family of bi-probabilities suffices to describe the perceptions of $S$-only observable $F$ for all non-classical observers. Therefore, we can say that the appearance of the quantum observable described by $\{Q_n^F\}_{n=1}^\infty$ is inter-subjective among non-classical observers.

Finally, since $P_n$'s are a (diagonal) part of $Q_n^F$'s, there is a commonality between the perceptions of classical and non-classical observers, making a comparison between them meaningful. The most obvious observation here is that, in general, the classical and non-classical perceptions differ due to the off-diagonal part of $Q_n^F$; in other words, the appearance of quantum observable $F$ is generally \textit{not} inter-subjective between classical and non-classical observers.

In closing, we wish to point out some notable connections between the bi-probability parameterization and other established theoretical approaches found in the literature. Inspection of Eq.~\eqref{eq:observer_state} suggests a possible intuitive interpretation for the components of bi-stochastic process $(F_+(t),F_-(t))$. The $F_+(t)$-component appears to play the role of an external field driving the ``branch'' of evolution going forwards in time, while $F_-(t)$ drives the backwards-in-time evolution branch. Therefore, the non-classical off-diagonal part of bi-probabilities would be responsible for the quantum interference between the forwards and backwards branches. The classical diagonal part $\delta_{\mathbf{f}_n,\mathbf{f}_{-n}}P_n(\mathbf{f}_n,\mathbf{t}_n)$ comes into play only when $F_-(t)$ overlaps with $F_+(t)$, and both components merge into a single field.

Such a structure of forwards and backwards fields is familiar from other approaches to open quantum systems, such as the Feynman-Vernon influence functional \cite{Feynman_AP63} or the Keldysh field theoretical approach \cite{Hofer_Quantum17,Chakraborty_PRB18,Wang_PRR20,Wang_NC21}. It should, however, be noted that components $F_+(t)$ and $F_-(t)$---the trajectories traced through the domain of the eigenvalues of operator $\hat F$---are fundamentally different objects compared to the paths of Feynman-Vernon approach and fermionic or bosonic fields in Keldysh formalism, as discussed in \cite{Szankowski_PRA21}. 

In contrast, the bi-probability parameterization of dynamics of observable $F$ can be directly linked with the \textit{consistent histories} formulation of quantum mechanics~\cite{Griffiths_JSP84,Gell-Mann_PRD93,Strasberg_SciPost23,Strasberg_arXiv23_2}. The link is straightforward: the central object of the histories formalism---the \textit{decoherence functional}---has the same form as the bi-probability, provided that the events constituting a history are identified with eigenspaces of observable $F$. The principal difference between the two lies in their respective origins. In the histories formalism, the decoherence functional is postulated as an improvement on the Born rule of the standard quantum theory; therefore, analogous to the Born rule, the decoherence functional has the status of primitive notion. Then, the diagonal part of the functional is used as a Born distribution while the off-diagonal part does not have an explicit utility. The bi-probabilities, on the other hand, are not postulated but rather identified as a facet of the standard formalism for describing the dynamics of quantum observable $F$. They appear naturally in the description of a quantum system $O$ coupled to $S$ through observable $F$: Eqs.~\eqref{eq:observer_state} and \eqref{eq:Q_quasi-avg_form} show that the reduced state of $O$, as well as any multi-time correlation functions of its observables, can all be expressed in terms of bi-average over the totality of bi-probabilities, including their off-diagonal parts. 

It could even be argued here that, through the link with bi-probabilities, the decoherence functional (or Born distribution, for that matter) can be seen as an emergent quantity derived from general formalism of standard quantum theory, rather than as a primitive notion as postulated in the original formulation of the consistent histories formalism.

\section{When the classical and non-classical observers are in agreement}\label{sec:KC_and_SF}

By combining the bi-consistency~\eqref{eq:quasi-KC} with the equation~\eqref{eq:Q_diag}, we establish the general relation between Born distributions mediated by the off-diagonal part of the bi-probabilities:
\begin{align}
\label{eq:generalized_KC}
        &P_{n-1}(f_n,t_n;\ldots;f_{i+1},t_{i+1};f_{i-1},t_{i-1}\ldots;f_1,t_1)-\sum_{f_i}P_n(\mathbf{f}_n,\mathbf{t}_n)
        = \prod_{j\neq i}\delta_{f_j,f_{-j}}\!\!\sum_{f_i \neq f_{-i}}\!\! Q^F_n(\mathbf{f}_n,\mathbf{f}_{-n},\mathbf{t}_n).
\end{align}
In the context of Kolmogorovian consistency (KC), the above equation shows that $\{P_n\}_{n=1}^\infty$ becomes consistent when the observable $F$ satisfies the \textit{consistent measurements} (CM) condition:
\begin{align}\label{eq:KCC_condition}
    \begin{aligned}[c]
        \text{\textbf{CM : }}&
    \end{aligned}
    &
    \begin{aligned}[c]
        &\forall(n\geqslant 1)\,\forall(0<t_1<\cdots<t_n)\,\forall(1\leqslant i \leqslant n)\,\forall(f_1,\ldots,f_{i-1},f_{i+1},\ldots,f_n):\\
        &0 =\prod_{j\neq i}\delta_{f_j,f_{-j}}\!\!\sum_{f_i\neq f_{-i}}\!\!Q^F_n(\mathbf{f}_n,\mathbf{f}_{-n},\mathbf{t}_n)\\
        &\phantom{0} = \sum_{f_i\neq f_{-i}}Q^F_n(f_n,f_{n},t_n;\ldots;f_i,f_{-i},t_i;\ldots;f_1,f_{1},t_1)\\
        &\phantom{0} = \prod_{j\neq i}\delta_{f_j,f_{-j}}\!\!\sum_{f_i> f_{-i}}\!\!2\operatorname{Re}\{Q^F_n(\mathbf{f}_n,\mathbf{f}_{-n},\mathbf{t}_n)\}.
    \end{aligned}
\end{align}
Thus, the fulfillment of CM (which is mathematically equivalent to KC) guarantees that the trajectory picture of $F$ applies to the observations made by the classical observer. However, not much can be said about the implications for contexts other than sequential measurements; in particular, it is unclear how it would affect the perceptions of non-classical observers. To address this issue, we propose considering a simpler but also stricter condition---we say that the observable $F$ satisfies the \textit{surrogate field} (SF) condition when
\begin{align}\label{eq:SFR_condition}
    \begin{aligned}[c]
        \text{\textbf{SF : }}&
    \end{aligned}
    &
    \begin{aligned}[c]
        &\forall(n \geqslant 1)\,\forall(\mathbf{f}_n,\mathbf{f}_{-n})\,\forall(0<t_1<\cdots<t_n):\\ 
        &\mathbf{f}_{-n} \neq \mathbf{f}_n\implies Q_n^F(\mathbf{f}_n,\mathbf{f}_{-n},\mathbf{t}_n) = 0
        \quad\iff\quad
        Q_n^F(\mathbf{f}_n,\mathbf{f}_{-n},\mathbf{t}_n) = \delta_{\mathbf{f}_n,\mathbf{f}_{-n}}P_n(\mathbf{f}_n,\mathbf{t}_n),    
    \end{aligned}
\end{align}
meaning the off-diagonal part of each bi-probability vanish, or equivalently, only the diagonal part of each bi-probability is non-zero. The SF (the reason for the name will soon become apparent) clearly implies CM\&KC, and more importantly, its broader physical implications for other contexts---especially for non-classical observers---are significantly more transparent. When the off-diagonal parts all vanish, bi-probabilities reduce to proper probabilities equal to Born distributions, which are consistent because CM\&KC are already implied. This means that, effectively, the two components of the bi-stochastic process $(F_+(t),F_-(t))$ in any bi-average [e.g., like in~\eqref{eq:observer_state}] are merged into a single proper stochastic process $F(t)$ defined by now consistent family $\{P_n\}_{n=1}^\infty$. In short, the bi-averages reduce to the standard stochastic averages over the trajectory distribution $P^F[f]$ combined from $P_n$'s. Consequently, the reduce density matrix of the non-classical observer simplifies to
\begin{align}
\nonumber
    \hat\varrho_o(t) & = \big\langle \hat V_o[F_+](t,0)\hat\rho_o\hat V_o[F_-](0,t)\big\rangle 
        = \iint Q^F[f_+,f_-]\hat V_o[f_+](t,0)\hat\rho_o\hat V_o[f_-](0,t)[Df_+][Df_-]\\
\label{eq:SFR}
    &\xrightarrow{\text{\textbf{SF}}} \int P^F[f]\hat V_o[f](t,0)\hat\rho_o\hat V_o[f](0,t)[Df] = \overline{\hat V_o[F](t,0)\hat\rho_o\hat V_o[F](0,t)},
\end{align}
since $Q_n^F(\mathbf{f}_n,\mathbf{f}_{-n},\mathbf{t}_n) = \delta_{\mathbf{f}_n,\mathbf{f}_{-n}}P_n(\mathbf{f}_n,\mathbf{t}_n)$ for all $n$ implies $Q^F[f_+,f_-]=\delta[f_+-f_-]P^F[f_+]$ in the bi-average notation. Therefore, when SF is satisfied, the dynamics in $O$ appear as if the system were driven by an external noise instead of being coupled to actual quantum system $S$. In other words, as far as the dynamics in $O$ are concerned, the coupling to the quantum observable operator $\hat F$ in $SO$ Hamiltonian~\eqref{eq:H_OS} has been replaced by an external \textit{surrogate field}~\cite{Szankowski_SciRep20} represented by the stochastic process $F(t)$,
\begin{align}
    &\hat H_{os} \to \hat H_o[F](t) = \hat H_o + \lambda F(t)\hat G_o,
\end{align}
and this stochastic Hamiltonian generates the evolution~\eqref{eq:SFR},
\begin{align}
   &\overline{Te^{-i\int_0^t\hat H_o[F](\tau)d\tau}\hat\rho_o \Big(Te^{-i\int_0^t\hat H_o[F](\tau)d\tau}\Big)^\dagger}
   = e^{-it\hat H_o}\overline{\hat V_o[F](t,0)\hat\rho_o\hat V_o[F](0,t)}e^{it\hat H_o}.
\end{align}

To summarize, we have identified here the surrogate field condition~\eqref{eq:SFR_condition} under which the sequential measurement of observable $F$ and the dynamics of a quantum system coupled to the operator representation $\hat F$ are both indistinguishable from their respective stochastic simulations. Moreover, crucially, the stochastic process used in these simulations in both contexts is the same surrogate field $F(t)$ defined by the family of bi-probabilities $\{Q_n^F\}_{n=1}^\infty$, which satisfy surrogate field condition and reduce to consistent Born distributions $\{P_n\}_{n=1}^\infty$. In other words, when the observable $F$ meets the surrogate field condition, the perceptions of this physical quantity by both classical and non-classical observers align. It must be underlined that this symmetry between classical and non-classical perceptions is not only a philosophical curiosity but also a physical effect with tangible consequences. 

For instance, consider a scenario where system $S$ acts as the environment and we aim to predict the evolution of the reduced state $\hat\varrho_o(t)$ of system $O$ coupled to $\hat F$. When the surrogate field condition is satisfied and stochastic simulation of the dynamics in $O$ is enabled, computing of the stochastic average in~\eqref{eq:SFR} poses the main challenge. This problem can be circumvented entirely if one also has access to the results of sequential measurements of $F$, which, in this case, are equivalent to sampling trajectories of the surrogate field $F(t)$. Indeed, since the measurement results and the surrogate field driving $O$ both are described by the same probability distributions $\{P_n\}_{n=1}^\infty$, the measured samples can be used to approximate the stochastic average with the sample average~\cite{Szankowski_PRA21},
\begin{align}
\nonumber
    \overline{\hat V_o[F](t,0)\hat\rho_o\hat V_o[F](0,t)} &= \int P^F[f]\hat V_o[f](t,0)\hat\rho_o\hat V_o[f](0,t)[Df]\\
\label{eq:approx_sample_avg}
    &\approx \frac{1}{N}\sum_{j=1}^N\hat V_o[f_j](t,0)\hat\rho_o\hat V_o[f_j](0,t),
\end{align}
where $\{f_j(t)\}_{j=1}^N$ is an ensemble of trajectories interpolated from the measured sequences [compare with Eq.~\eqref{eq:stochastic_avg}]. Equation~\eqref{eq:approx_sample_avg} explicitly demonstrates that the trajectory picture emerging for SF-satisfying observable applies equally in both measurement and non-measurement contexts---it shows that the trajectories are interchangeable between these contexts. 

Therefore, when evaluated against our standard of objectivity, the trajectory picture is promoted from being inter-subjective solely among classical observers to being inter-subjective between classical and non-classical observers. However, since classical and non-classical observers (as well as measurement and non-measurement contexts) are categorically distinct on a fundamental level in quantum theory, it is inadequate to simply assert that the trajectory picture of SF observable is ``more'' inter-subjective than, for example, non-SF trajectories perceived only by classical observers. Such a cross-category extension of inter-subjectivity---especially when the encompassed categories are so incompatible from any other point of view---demands a categorical rather than merely quantitative extension of the notion itself. Of course, \textit{objectivity} is the natural extension for inter-subjectivity that connects almost all (if not all) possible types of observers~\cite{Nagel_book}, providing justification for our original claim.

Returning to the specific question of Kolmogorov consistency (KC) of Born distributions, we make a final note on the relationship between the consistent measurements (CM) condition and the surrogate field (SF) condition, when they are treated only as the enablers of KC. From a purely mathematical perspective, the CM and SF conditions are evidently not equivalent, at least when considering the bi-probabilities $Q_n^F$ in isolation. If $\{Q_n^F\}_{n=1}^\infty$ were arbitrary functions, the most that could be stated is that CM is a weaker condition than SF (or conversely, SF is stronger than CM), since SF implies CM but not vice versa. However, $Q_n^F$'s are not arbitrary; to the contrary, each bi-probability possesses a complex internal structure arising from interactions between the dynamical law $\hat U$, the partitioning $\{\hat P(f)\}_f$, and the initial condition $\hat\rho$. How this internal complexity affects the interrelations between SF and CM is, predictably, an exceedingly difficult problem to solve exactly. 

In such situations, when the mathematically strict answer is not forthcoming, the relative significance of conditions can be estimated based on empirical evidence from a variety of cases. If no physical system is known that satisfies the formally weaker CM without also satisfying the formally stronger SF, the likelihood that CM is spurious (i.e., satisfying CM requires first satisfying SF) increases. Indeed, to our best knowledge, this appears to be the case. On one hand, numerous examples of physical systems satisfying SF is readily identifiable; some are discussed Appendix~\ref{sec:SFR_examples}, and further examples can be found in~\cite{Strasberg_arXiv23_2}. On the other hand, we are not aware of any system that satisfies CM alone without also satisfying SF, strongly suggesting that the consistent measurement condition may lack physical significance. Unlike CM, which is relevant exclusively in the context of sequential measurements, the surrogate field condition warrants further consideration due to its broad implications across diverse contexts beyond Kolmogorovian consistency. These include surrogate field representations, perceptions by non-classical observers, and dynamics of open systems in general. In addition to the examples discussed here, consistent histories theory should also be noted, as the SF condition is formally equivalent to the \textit{decoherent histories} condition~\cite{Diosi_PRL04,Strasberg_arXiv23_2}, provided that one identifies the decoherence functional as the bi-probability.

\section{Conclusions}

We have identified the condition---the surrogate field (SF) criterion~\eqref{eq:SFR_condition}---under which a quantum observable $F$ appears to both classical and non-classical observers as a stochastic process $F(t)$. For a classical observer, this manifests as consistent measurements of $F$, i.e.~sequential measurements described by the Born probability distributions satisfying the Kolmogorov consistency (KC) criterion~\eqref{eq:Born_KC}. Consequently, the result of a measurement performed by the observer is equivalent to sampling a trajectory $f(t)$ that realizes the process $F(t)$. For non-classical observer---that is, a quantum system coupled to $F$---the system's dynamics can be fully described by a stochastic simulation where the quantum operator $\hat F$ representing the observable is replaced by the process $F(t)$. Crucially, this surrogate field, which mimics the coupling with an actual quantum system, is precisely the stochastic process sampled by the classical observer. Thus, when observable $F$ satisfies the SF condition, not only the quantum stochastic process describing the sequential measurement becomes classical, but it can also be considered an \textit{objective} entity.

Besides identifying the SF condition, we have also found a weaker consistent measurements (CM) condition~\eqref{eq:KCC_condition} for Kolmogorovian consistency of $\{P_n\}_{n=1}^\infty$. In fact, CM and KC conditions are equivalent, as directly implied by relation~\eqref{eq:generalized_KC} between Born distributions, which is a novel and interesting formal result in itself. However, we have argued that the stronger SF condition holds greater physical significance: when satisfied, it not only implies CM\&KC but, more importantly, it also brings about the symmetry between the perceptions of classical and non-classical observers. Hence, it seems more vital---even from practical point of view because of potential implementation of stochastic simulations of open system dynamics, see Eq.~\eqref{eq:approx_sample_avg}---to verify whether SF is satisfied, over merely confirming CM\&KC. If one accepts this assessment, then CM\&KC's primary value would lie in its potential ability to determine the status of SF. Even though is it likely that CM is actually a spurious condition (i.e., it can only be satisfied when SF is satisfied), we cannot dismiss the possibility that Kolmogorovian consistency might occur independently of SF. Thus, at present, the consistency (or rather its violation) can serve only as a witness of surrogate field condition violation.

\acknowledgements
The authors would like to thank D. Chru\'{s}ci\'{n}ski, D. Lonigro, F. Sakuldee, M. Gajda, F. Gampel, P. Strasberg, and S. Milz for many illuminating discussions and useful comments on the manuscript.

This work was supported by the Polish National Science Centre through the project MAQS under QuantERA, which has received funding from the European Union’s Horizon 2020 research and innovation program under Grant Agreement No. 731473, Project No. 2019/32/Z/ST2/00016.

\appendix
\section{Derivation of bi-average forms}\label{appx:quasi-average_form}
\subsection{Reduced state}
The objective is to show that the interaction picture of the reduced state of $O$,
\begin{align}
    \hat\varrho_o(t) &= \operatorname{tr}_s\left[\hat V_{os}(t,0)\hat\rho_o\otimes\hat\rho\hat V_{os}(0,t)\right],
\end{align}
where
\begin{align}
    \hat V_{os}(t,s) &= Te^{-i\lambda\int_s^t \hat G_o(\tau)\otimes\hat F(\tau)d\tau};\ 
    \hat G_o(\tau) = \hat U_o^\dagger(\tau)\hat G_o\hat U_o(\tau);\  \hat F(\tau) = \hat U^\dagger(\tau)\hat F\hat U(\tau),
\end{align}
can be rewritten in the bi-average form~\eqref{eq:observer_state}.

We begin by switching to the super-operator language (i.e., linear operators that act on operators):
\begin{align}
    \hat\varrho_o(t) &\equiv \Lambda_t\hat\rho_o,
\end{align}
where the dynamical map $\Lambda_t$ is a super-operator acting on operators in $O$ and is defined as
\begin{align}
\nonumber
    \Lambda_t &= \operatorname{tr}_s\left[\hat V_{os}(t,0)(\bullet\otimes\hat\rho)\hat V_{os}(0,t)\right]
        = \operatorname{tr}_s\left[Te^{-i\lambda\int_0^t [\hat G_o(\tau)\otimes\hat F(\tau),\bullet]d\tau}(\bullet\otimes\hat\rho)\right]\\
\nonumber
    &= \sum_{k=0}^\infty (-i\lambda)^k \int\limits_0^t\!\!d\tau_k\cdots\!\!\int\limits_0^{\tau_2}\!\!d\tau_1
        \operatorname{tr}_s\left([\hat G_o(\tau_k)\otimes\hat F(\tau_k),\cdots [\hat G_o(\tau_1)\otimes\hat F(\tau_1),\bullet\otimes\hat\rho]\cdots ]\right)\\
    &= \sum_{k=0}^\infty (-i\lambda)^k \int\limits_0^t\!\!d\tau_k\cdots\!\!\int\limits_0^{\tau_2}\!\!d\tau_1
        \operatorname{tr}_s\left([\hat G_o(\tau_k)\otimes\hat F(\tau_k),\bullet]\cdots[\hat G_o(\tau_1)\otimes\hat F(\tau_1),\bullet]({\bullet}\otimes\hat\rho)\right),
\end{align}
where the symbol $\bullet$ is understood as a ``placeholder'' for the argument of the super-operator acting on operators to its right, e.g., $[\hat A,\bullet]\hat B = [\hat A,\hat B]$, or $\bullet\hat A = \hat A$, etc. To arrive at this form we have made use of a well known identity for exponentials (time-ordered and otherwise),
\begin{align}
    Te^{-i\int_s^t \hat A(\tau)d\tau}\bullet\Big(Te^{-i\int_s^t\hat A(\tau)d\tau}\Big)^\dagger &= Te^{-i\int_s^t [\hat A(\tau),\bullet]d\tau}.
\end{align}

Next, we aim to rewrite the map as a moment series. First, we use the spectral decomposition of the observable $\hat F$,
\begin{align}
    \hat F(\tau) = \hat U^\dagger(\tau)\Big(\sum_f f\hat P(f)\Big)\hat U(\tau) = \sum_f f \hat P(f,\tau),
\end{align}
and we substitute it into the formula,
\begin{align}
\nonumber
    \Lambda_t &= \sum_{k=0}^\infty (-i\lambda)^k \int\limits_0^t\!\!d\tau_k\cdots\!\!\int\limits_0^{\tau_2}\!\!d\tau_1\\
    &\phantom{=}\times
        \operatorname{tr}_s\left(\sum_{f_k}f_k [\hat G_o(\tau_k)\otimes\hat P(f_k,\tau_k),\bullet]\cdots\sum_{f_1}f_1[\hat G_o(\tau_1)\otimes\hat P(f_1,\tau_1),\bullet]({\bullet}\otimes\hat\rho)\right).
\end{align}
Second, we show that the $S$ and $O$ parts of the above expression factorize; to see this, consider one of the super-operators $\sum_f f[\hat G_o(\tau)\otimes\hat P(f,\tau),\bullet]$ and act with it on an operator in an outer product form $\hat O\otimes\hat S$,
\begin{align}
\nonumber
    &\sum_{f}f[\hat G_o(\tau)\otimes\hat P(f,\tau),\bullet]\hat O\otimes\hat S\\ 
\nonumber
    &\phantom{=}
    = \sum_{f_+}f_+\big(\hat G_o(\tau)\hat O\big)\otimes\big(\hat P(f_+,\tau)\hat S\big) - \sum_{f_-}f_-\big(\hat O\hat G_o(\tau)\big)\otimes\big(\hat S\hat P(f_-,\tau)\big)\\
\nonumber
    &\phantom{=}
    = \sum_{f_+,f_-}f_+\big(\hat G_o(\tau)\hat O\big)\otimes\big(\hat P(f_+,\tau)\hat S \hat P(f_-,\tau)\big)
        - \sum_{f_+,f_-}f_-\big(\hat O\hat G_o(\tau)\big)\otimes\big(\hat P(f_+,\tau)\hat S\hat P(f_-,\tau)\big)\\
\nonumber
    &\phantom{=}
    = \sum_{f_+,f_-}\big(f_+\hat G_o\hat O - \hat O\hat G_o(\tau)f_-\big)\otimes\big(\hat P(f_+,\tau)\hat S\hat P(f_-,\tau)\big)\\
\nonumber
    &\phantom{=}
    = \sum_{f_+,f_-}\big(f_+\hat G_o(\tau){\bullet}-{\bullet}\hat G_o(\tau)f_-\big)\otimes\big(\hat P(f_+,\tau){\bullet}\hat P(f_-,\tau)\big)(\hat O\otimes\hat S)\\
\label{eq::appx::sep_for_product}
    &\phantom{=}\equiv \sum_{f_+,f_-}\mathcal{G}_{\tau}(f_+,f_-)\otimes\mathcal{P}_\tau(f_+,f_-)(\hat O\otimes\hat S),
\end{align}
which can then be easily generalized to the super-operator equality,
\begin{align}\label{eq:appx:sep_gen}
    \sum_f f[\hat G_o(\tau)\otimes\hat P(f,\tau),\bullet] &= \sum_{f_+,f_-}\mathcal{G}_\tau(f_+,f_-)\otimes\mathcal{P}_\tau(f_+,f_-).
\end{align}
To see this, take a product basis in the $O+S$ operator space equipped with a standard scalar product $(A|B) = \operatorname{tr}_{os}(\hat A^\dagger\hat B)$: if $\{\hat O_i\}_i$ is an orthonormal basis in $O$-operator space, and $\{\hat S_\alpha\}_\alpha$ is a basis in $S$, then $\{\hat O_i\otimes\hat S_\alpha\}_{i,\alpha}$ is a basis in $O{+}S$-operator space. With this basis, any arbitrary operator $\hat A_{os}$ can be decomposed into linear combination of factorized products, $\hat A_{os} = \sum_{i,\alpha}A_{i\alpha}\hat O_i\otimes\hat S_\alpha$; the general result~\eqref{eq:appx:sep_gen}, then follows directly from the special result~\eqref{eq::appx::sep_for_product}. 

Thanks to the factorization of $S$ and $O$ parts, we arrive at the form where the partial trace over $S$ can be computed explicitly,
\begin{align}
\nonumber
    \Lambda_t &= \sum_{k=0}^\infty (-i\lambda)^k \int\limits_0^t\!\!d\tau_k\cdots\!\!\int\limits_0^{\tau_2}\!\!d\tau_1\sum_{f_k,f_{-k}}\cdots\sum_{f_1,f_{-1}}\\
\nonumber
    &\phantom{=}\times
        \operatorname{tr}_s\left[\mathcal{P}_{\tau_k}(f_k,f_{-k})\cdots\mathcal{P}_{\tau_1}(f_1,f_{-1})\hat\rho\right]
        \mathcal{G}_{\tau_k}(f_k,f_{-k})\cdots\mathcal{G}_{\tau_1}(f_1,f_{-1})\\
\nonumber
    &=\sum_{k=0}^\infty (-i\lambda)^k \int\limits_0^t\!\!d\tau_k\cdots\!\!\int\limits_0^{\tau_2}\!\!d\tau_1\sum_{f_k,f_{-k}}\cdots\sum_{f_1,f_{-1}}\\
\nonumber
    &\phantom{=}\times
        \operatorname{tr}_s\Big[\hat P(f_k,\tau_k)\cdots\hat P(f_1,\tau_1)\hat\rho\hat P(f_{-1},\tau_1)\cdots\hat P(f_{-k},\tau_k)\Big]
        \mathcal{G}_{\tau_k}(f_k,f_{-k})\cdots\mathcal{G}_{\tau_1}(f_1,f_{-1})\\
    &=\sum_{k=0}^\infty (-i\lambda)^k \int\limits_0^t\!\!d\tau_k\cdots\!\!\int\limits_0^{\tau_2}\!\!d\tau_1\sum_{f_k,f_{-k}}\cdots\sum_{f_1,f_{-1}}
        Q_k^F(\mathbf{f}_k,\mathbf{f}_{-k},\bm{\tau}_k)\mathcal{G}_{\tau_k}(f_k,f_{-k})\cdots\mathcal{G}_{\tau_1}(f_1,f_{-1}).
\end{align}
We have recognized in the traced factor the formula for the bi-probability~\eqref{eq:Q}. We are also recognizing here a bi-average in each term of the series,
\begin{align}
\nonumber
    &\sum_{f_k,f_{-k}}\cdots\sum_{f_1,f_{-1}} Q_k^F(\mathbf{f}_k,\mathbf{f}_{-k},\bm{\tau}_k)\mathcal{G}_{\tau_k}(f_k,f_{-k})\cdots\mathcal{G}_{\tau_1}(f_1,f_{-1})\\
\nonumber
    &\phantom{=}=
        \sum_{f_k,f_{-k}}\cdots\sum_{f_1,f_{-1}}\iint [Df_+][Df_-] Q^F[f_+,f_-]\,\Big(\prod_{i=1}^k\delta(f_i - f_+(\tau_i))\delta(f_{-i} - f_-(\tau_i)\Big)\\
\nonumber
    &\phantom{==}\times\mathcal{G}_{\tau_k}(f_+(\tau_k),f_-(\tau_k))\cdots\mathcal{G}_{\tau_1}(f_+(\tau_1),f_-(\tau_1))\\
\nonumber
    &\phantom{=}=\iint Q^F[f_+,f_-]\mathcal{G}_{\tau_k}(f_+(\tau_k),f_-(\tau_k))\cdots\mathcal{G}_{\tau_1}(f_+(\tau_1),f_-(\tau_1))[Df_+][Df_-]\\
    &\phantom{=}=\left\langle \mathcal{G}_{\tau_k}(F_+(\tau_k),F_-(\tau_k))\cdots\mathcal{G}_{\tau_1}(F_+(\tau_1),F_-(\tau_1)) \right\rangle.
\end{align}
As a result, we can take the whole series under the sign of bi-average,
\begin{align}
\nonumber
    \Lambda_t &= \sum_{k=0}^\infty (-i\lambda)^k \int\limits_0^t\!\!d\tau_k\cdots\!\!\int\limits_0^{\tau_2}\!\!d\tau_1
        \left\langle \mathcal{G}_{\tau_k}(F_+(\tau_k),F_-(\tau_k))\cdots\mathcal{G}_{\tau_1}(F_+(\tau_1),F_-(\tau_1)) \right\rangle\\
\nonumber
    &= \left\langle \sum_{k=0}^\infty (-i\lambda)^k \int\limits_0^t\!\!d\tau_k\cdots\!\!\int\limits_0^{\tau_2}\!\!d\tau_1 
        \mathcal{G}_{\tau_k}(F_+(\tau_k),F_-(\tau_k))\cdots\mathcal{G}_{\tau_1}(F_+(\tau_1),F_-(\tau_1)) \right\rangle\\
\label{eq:appx:exp_of_Gamma}
    &= \Big\langle Te^{-i\lambda\int_0^t \mathcal{G}_\tau(F_+(\tau),F_-(\tau))d\tau}\Big\rangle.
\end{align}

The last step is to demonstrate that the series under the bi-average evaluates to the form presented in Eq.~\eqref{eq:observer_state}, i.e., we must prove that
\begin{align}
    Te^{-i\lambda\int_0^t\mathcal{G}_\tau(F_+(\tau),F_-(\tau))d\tau} &= \hat V_{o}[F_+](t,0)\bullet\hat V_{o}[F_-](0,t).
\end{align}
The simplest way to show it exploits the fact that the two super-operators constituting $\mathcal{G}_\tau$,
\begin{align}
    \mathcal{G}_\tau(F_+(\tau),F_-(\tau)) = F_+(\tau)\hat G_o(\tau){\bullet} + (-F_-(\tau){\bullet}\hat G_o(\tau)) \equiv \mathcal{G}_+(\tau) + \mathcal{G}_{-}(\tau),
\end{align}
commute with each other,
\begin{align}
\nonumber
    \forall(\tau,\tau'): 
        \mathcal{G}_+(\tau)\mathcal{G}_-(\tau') &= -F_+(\tau)F_-(\tau')
        (\hat G_o(\tau){\bullet})({\bullet}\hat G_o(\tau'))
        = -F_+(\tau)F_-(\tau')\hat G_o(\tau){\bullet}\hat G_o(\tau')\\
        &=-F_+(\tau)F_-(\tau')({\bullet}\hat G_o(\tau'))(\hat G_o(\tau){\bullet}) = \mathcal{G}_-(\tau')\mathcal{G}_+(\tau).
\end{align}
If so, then the exponential function in~\eqref{eq:appx:exp_of_Gamma} simply factorizes into composition of two (commuting) exponentials,
\begin{align}
\nonumber
    Te^{-i\lambda\int_0^t\mathcal{G}_\tau(F_+(\tau),F_-(\tau))d\tau} &= 
        \left(Te^{-i\lambda\int_0^t\mathcal{G}_{+}(\tau)d\tau}\right)\left(Te^{-i\lambda\int_0^t\mathcal{G}_{-}(\tau)d\tau}\right)\\
\nonumber
    &= Te^{\left(-i\lambda\int_0^t F_+(\tau)\hat G_o(\tau)d\tau\right){\bullet}}\ Te^{{\bullet}\left({+}i\lambda\int_0^t F_-(\tau)\hat G_o(\tau)d\tau\right)}\\
\nonumber
    &= \left(\sum_{k=0}^\infty(-i\lambda)^k\int\limits_0^t\!\!d\tau_k\cdots\!\!\int\limits_0^{\tau_2}\!\!d\tau_1 
        F_+(\tau_k)\hat G_o(\tau_k)\cdots F_+(\tau_1)\hat G_o(\tau_1)\right){\bullet}\\
\nonumber
    &\phantom{=}\times
        {\bullet}\left(\sum_{k=0}^\infty (i\lambda)^k\int\limits_0^t\!\!d\tau_k\cdots\!\!\int\limits_0^{\tau_2}\!\!d\tau_1
        F_-(\tau_1)\hat G_o(\tau_1)\cdots F_-(\tau_k)\hat G_o(\tau_k)\right)\\
\label{eq:appx:propagator_identity}
    &=\hat V_o[F_+](t,0){\bullet}\big(\hat V_o[F_-](t,0)\big)^\dagger = \hat V_o[F_+](t,0){\bullet}\hat V_o[F_-](0,t).
\end{align}

\subsection{Bi-probabilities}
Here, the goal is to show that the general formula~\eqref{eq:Q} for bi-probability associated with the $O$-only observable, say $\hat X_o= \sum_x x \hat P_o(x)$,
\begin{align}
    Q_n^{X_o}(\mathbf{x}_n,\mathbf{x}_{-n},\mathbf{t}_n) = \operatorname{tr}\left[
        \left(\prod_{i=n}^1 \hat U_{os}^\dagger(t_i)\hat P_o(x_i)\otimes\hat 1 \hat U_{os}(t_i)\right)
        \hat\rho_o\otimes\hat\rho
        \left(\prod_{i=1}^n\hat U_{os}^\dagger(t_i)\hat P_o(x_{-i})\otimes\hat 1\hat U_{os}(t_i)\right)\right],
\end{align}
can also be written in the bi-average form.

First, we translate the general definition of bi-probability in the language of super-operators,
\begin{align}
\nonumber
    &Q_n^{X_o}(\mathbf{x}_n,\mathbf{x}_{-n},\mathbf{t}_n)\\
\nonumber
    &\phantom{=}
    = \operatorname{tr}\left[
        \left(\prod_{i=n}^1 \hat V_{os}(0,t_i)\big(\hat U_o^\dagger(t_i)\hat P_o(x_i)\hat U_o(t_i)\big)\otimes\big(\hat U^\dagger(t_i)\hat U(t_i)\big)\hat V_{os}(t_i,0)\right)\hat\rho_o\otimes\hat\rho
        \Bigg(\cdots\Bigg)
    \right]\\
\nonumber
    &\phantom{=}
    = \operatorname{tr}\left[
        \left(\prod_{i=n}^1 \big(\hat P_o(x_i,t_i)\otimes\hat 1\big)\hat V_{os}(t_{i},t_{i-1})\right)\hat\rho_o\otimes\hat\rho
        \left(\prod_{i=1}^{n}\hat V_{os}(t_{i-1},t_i)\big(\hat P_o(x_{-i},t_i)\otimes\hat 1\big)\right)
    \right]\\
\nonumber
    &\phantom{=}
    = \operatorname{tr}\left[
        \big(\hat P_o(x_n,t_n)\otimes \hat 1 \bullet \hat P_o(x_{-n},t_n)\otimes\hat 1\big)\big(\hat V_{os}(t_n,t_{n-1}){\bullet}\hat V_{os}(t_{n-1},t_n)\big)\cdots\right.\\
\nonumber
    &\phantom{==\operatorname{tr}[}\left.
    \cdots\big(\hat P_o(x_1,t_1)\otimes\hat 1 \bullet \hat P_o(x_{-1},t_1)\otimes\hat 1\big)\big(\hat V_{os}(t_1,0){\bullet}\hat V_{os}(0,t_1)\big)\hat\rho_o\otimes\hat\rho
    \,\right]\\
    &\phantom{=}
    \equiv \operatorname{tr}\left[
        \big(\mathcal{P}_{t_n}(x_n,x_{-n})\otimes{\bullet}\big)\Omega_{os}(t_n,t_{n-1})\cdots\big(\mathcal{P}_{t_1}(x_1,x_{-1})\otimes{\bullet}\big)\Omega_{os}(t_1,0)\hat\rho_o\otimes\hat\rho
    \,\right],
\end{align}
where $t_{0} = 0$ and we have used the definition of the interaction picture of evolution operator~\eqref{eq:U_os}.

Second, using methods (and notation) from previous section, we rewrite the propagators $\Omega$'s in the terms of generators $\mathcal{G}$'s,
\begin{align}
    \Omega_{os}(t,s) &= \sum_{k=0}^\infty (-i\lambda)^k\int\limits_s^t\!\!d\tau_k\cdots\!\!\int\limits_s^{\tau_2}\!\!d\tau_1
        \sum_{\mathbf{f}_k,\mathbf{f}_{-k}}\left(\prod_{i=k}^1\mathcal{G}_{\tau_k}(f_i,f_{-i})\right)
        \otimes\left(\prod_{i=k}^1\mathcal{P}_{\tau_k}(f_i,f_{-i})\right).
\end{align}
When we substitute this form into $Q_n^{X_o}$ we can, again, compute the trace over $S$ explicitly,
\begin{align}
\nonumber
    &Q_n^{X_o}(\mathbf{x}_n,\mathbf{x}_{-n},\mathbf{t}_n)\\
\nonumber
    &\phantom{=}
    = \sum_{k_n = 0}^\infty\cdots\sum_{k_1 = 0}^\infty (-i\lambda)^{\sum_{i=1}^n k_i}
        \left(\int\limits^{t_n}_{t_{n-1}}\!\!d\tau_{k_n}^n\cdots\!\!\int\limits^{\tau_{2}^n}_{t_{n-1}}\!\!d\tau_1^n\right)
        \cdots\left(\int\limits_0^{t_1}\!\!d\tau_{k_1}^1\cdots\!\!\int\limits_0^{\tau_{2}^1}\!\!d\tau_1^1\right)
        \sum_{\mathbf{f}_{k_n}^n}\sum_{\mathbf{f}_{-k_n}^n}\cdots\sum_{\mathbf{f}_{k_1}^1}\sum_{\mathbf{f}_{-k_1}^1}\\
\nonumber
    &\phantom{===}\times
        \operatorname{tr}_s\left[\mathcal{P}_{\tau_{k_n}^n}(f_{k_n}^n,f_{-k_n}^n)\cdots\mathcal{P}_{\tau_{1}^1}(f_1^1,f_{-1}^1)\hat\rho\right]\\
\nonumber
    &\phantom{====}\times
        \operatorname{tr}_o\left[
            \mathcal{P}_{t_n}(x_n,x_{-n})\mathcal{G}_{\tau_{k_n}^n}(f_{k_n}^n,f_{-k_n}^n)\cdots\mathcal{G}_{\tau_1^n}(f_1^n,f_{-1}^n)\cdots\right.\\
\nonumber
    &\phantom{====\times\operatorname{tr}_o[}\left.
            \cdots\mathcal{P}_{t_1}(x_1,x_{-1})\mathcal{G}_{\tau_{k_1}^1}(f_{k_1}^1,f_{-k_1}^1)\cdots\mathcal{G}_{\tau_1^1}(f_1^1,f_{-1}^1)\hat\rho_o
        \right]\\
\nonumber
    &\phantom{=}
     = \sum_{k_n = 0}^\infty\cdots\sum_{k_1 = 0}^\infty (-i\lambda)^{\sum_{i=1}^n k_i}
        \left(\int\limits^{t_n}_{t_{n-1}}\!\!d\tau_{k_n}^n\cdots\!\!\int\limits^{\tau_{2}^n}_{t_{n-1}}\!\!d\tau_1^n\right)
        \cdots\left(\int\limits_0^{t_1}\!\!d\tau_{k_1}^1\cdots\!\!\int\limits_0^{\tau_{2}^1}\!\!d\tau_1^1\right)
        \sum_{\mathbf{f}_{k_n}^n}\sum_{\mathbf{f}_{-k_n}^n}\cdots\sum_{\mathbf{f}_{k_1}^1}\sum_{\mathbf{f}_{-k_1}^1}\\
\nonumber
    &\phantom{===}\times
    Q_{k_n+\cdots+k_1}^F(f_{k_n}^n,f_{-k_n}^n,\tau_{k_n}^n;\ldots;f_{1}^1,f_{-1}^1,\tau_1^1)\\
\nonumber
    &\phantom{====}\times
    \operatorname{tr}_o\left[
        \mathcal{P}_{t_n}(x_n,x_{-n})\mathcal{G}_{\tau_{k_n}^n}(f_{k_n}^n,f_{-k_n}^n)\cdots\mathcal{G}_{\tau_1^1}(f_1^1,f_{-1}^1)\hat\rho_o
    \right]\\
\nonumber
    &\phantom{=}
     = \operatorname{tr}_o\Big[\Big\langle \mathcal{P}_{t_n}(x_n,x_{-n})
        \Bigg(\!\sum_{k_n = 0}^\infty(-i\lambda)^{k_n}\!\!\int\limits^{t_n}_{t_{n-1}}\!\!\!d\tau_{k_n}^n\cdots\!\!\!\int\limits^{\tau_{2}^n}_{t_{n-1}}\!\!\!d\tau_1^n
            \prod_{i=k_n}^1\mathcal{G}_{\tau_{i}^n}(F_+(\tau_{i}^n),F_-(\tau_{i}^n))
        \!\Bigg)\\
\nonumber
    &\phantom{====}\cdots
        \mathcal{P}_{t_1}(x_1,x_{-1})
        \Bigg(\sum_{k_1 = 0}^\infty(-i\lambda)^{k_1}\!\!\int\limits^{t_1}_{0}\!\!d\tau_{k_1}^1\cdots\!\!\int\limits^{\tau_{2}^1}_{0}\!\!d\tau_1^1
            \prod_{i=k_1}^1\mathcal{G}_{\tau_{i}^1}(F_+(\tau_{i}^1),F_-(\tau_{i}^1))
        \Bigg)\Big\rangle\hat\rho_o
    \Big]\\
\label{eq:appx:Q_obs_super-prop_form}
    &\phantom{=}=
    \operatorname{tr}\left[\Big\langle
        \mathcal{P}_{t_n}(x_n,x_{-n}) Te^{-i\lambda\int_{t_{n-1}}^{t_n}\mathcal{G}_\tau(F_+(\tau),F_-(\tau))d\tau}
        \cdots\mathcal{P}_{t_1}(x_1,x_{-1})Te^{-i\lambda\int_0^{t_1}\mathcal{G}_\tau(F_+(\tau),F_-(\tau))d\tau}\Big\rangle
        \hat\rho_o
    \right].
\end{align}
Where we have again recognized the definition of $Q_n^F$ in the term traced over $S$. The final step is to use the identity~\eqref{eq:appx:propagator_identity},
\begin{align}
\nonumber
    &Q_n^{X_o}(\mathbf{x}_n,\mathbf{x}_{-n},\mathbf{t}_n)\\
\nonumber
    &\phantom{=}=
    \operatorname{tr}\left[\left\langle
        \big(\hat P_o(x_n,t_n){\bullet}\hat P_o(x_{-n},t_n)\big)\big(\hat V_o[F_+](t_n,t_{n-1}){\bullet}\hat V_o[F_-](t_{n-1},t_n)\big)\cdots\right.\right.\\
\nonumber
    &\phantom{==\operatorname{tr}\langle}\left.\left.
        \cdots\big(\hat P_o(x_1,t_1){\bullet}\hat P_o(x_{-1},t_1)\big)\big(\hat V_o[F_+](t_1,0){\bullet}\hat V_o[F_-](0,t_1)\big)
    \right\rangle\hat\rho_o\right]\\
\nonumber
    &\phantom{=}=
    \operatorname{tr}\left\langle
        \left(\prod_{i=n}^1 \hat P_o(x_i,t_i)\hat V_o[F_+](t_i,t_{i-1})\right)\hat\rho_o\left(\prod_{i=1}^n\hat V_o[F_-](t_{i-1},t_i)\hat P_o(x_{-i},t_i)\right)
    \right\rangle\\
    &\phantom{=}=
    \operatorname{tr}\left\langle
        \left(\prod_{i=n}^1 \hat V_o[F_+](0,t_i)\hat P_o(x_i,t_i)\hat V_o[F_+](t_i,0)\right)\hat\rho_o
        \left(\prod_{i=n}^1 \hat V_o[F_-](0,t_i)\hat P_o(x_{-i},t_i)\hat V_o[F_-](t_i,0)\right)
    \right\rangle.
\end{align}

\section{Examples of bi-probabilities}\label{sec:SFR_examples}
\subsection{Quasi-static observable}
As the first example consider the case when
\begin{align}
    [\hat F, \hat H] = 0,
\end{align}
so that the Heisenberg picture of the observable is static, $\hat F(t) = \hat F$ [and thus, $\hat P(f,s) = \hat P(f)$]. Then, the bi-probabilities have a very simple form,
\begin{align}
\nonumber
    Q_n^F(\mathbf{f}_n,\mathbf{f}_{-n},\mathbf{t}_n) &= \operatorname{tr}\left[\left(\prod_{i=n}^1\hat P(f_i,t_i)\right)\hat\rho\left(\prod_{i=1}^n\hat P(f_{-i},t_i)\right)\right]
        =\operatorname{tr}\left[\left(\prod_{i=n}^1\hat P(f_i)\right)\hat\rho\left(\prod_{i=1}^n\hat P(f_{-i})\right)\right]\\
    &= \delta_{f_n,f_{-n}}\Big(\prod_{i=1}^{n-1} \delta_{f_n,f_{i}}\delta_{f_{-n},f_{-i}}\Big)\operatorname{tr}\big[\hat P(f_n)\hat\rho\big]
        = \delta_{\mathbf{f}_n,\mathbf{f}_{-n}}\Big(\prod_{i=1}^{n-1} \delta_{f_n,f_i}\Big)\operatorname{tr}\big[\hat P(f_n)\hat\rho\big].
\end{align}
As a result, the off-diagonal part vanishes automatically, and thus, the bi-stochastic process $(F_+(t),F_-(t))$ is reduced to a single-component time-independent stochastic variable $F$ with probability distribution $P^F(f) = \operatorname{tr}[\hat P(f)\hat\rho]$. Therefore, the system satisfies the surrogate field (SF) condition~\eqref{eq:SFR_condition} with a constant surrogate field.

\subsection{Pointer observable}
Consider a system $S$ composed of two interacting subsystems, $A$ and $B$,
\begin{align}
    \hat H = \hat H_{ab} = \hat H_a \otimes \hat 1 + \hat 1\otimes \hat H_b + \mu \hat G_a\otimes\hat G_b;\quad \hat\rho = \hat\rho_a\otimes\hat\rho_b,
\end{align}
but the observable $F$ belongs only to $A$,
\begin{align}
    \hat F = \sum_f f \hat P_a(f)\otimes \hat 1.
\end{align}
In words: $A$ can be treated as a non-classical observer of $\hat G_b$, or, equivalently, $B$ can be thought of as an environment to the open system $A$. In that case, we can use the result of appendix~\eqref{appx:quasi-average_form} to express the bi-probabilities associated with $F$ in terms of bi-average over $Q^{G_b}[g_+,g_-]$---bi-probabilities associated with observable $G_b$ of subsystem $B$,
\begin{align}\label{eq:appx:Q_open_env}
    Q_n^F(\mathbf{f}_n,\mathbf{f}_{-n},\mathbf{t}_n) &= \operatorname{tr}_a\left[ 
        \left\langle\prod_{i=n}^1\mathcal{P}_{t_i}(f_i,f_{-i})\Omega[G_+,G_-](t_i,t_{i-1})\right\rangle\hat\rho_a
    \right],
\end{align}
where $t_{0} = 0$, the symbol $\prod_{i=n}^1 \Lambda_i$ indicates an ordered composition of super-operators, $\Lambda_n\cdots\Lambda_1$, and
\begin{subequations}
\begin{align}
    \mathcal{P}_t(f_+,f_-) &= \hat P_a(f_+,t){\bullet}\hat P_a(f_-,t) = e^{it\hat H_a}\hat P_a(f_+)e^{-it\hat H_a}{\bullet}e^{it\hat H_a}\hat P_a(f_-)e^{-it\hat H_a};\\
    \Omega[g_+,g_-](t,s) &= Te^{-i\mu\int_s^t \mathcal{G}_\tau(g_+(\tau),g_-(\tau))d\tau};\\
    \mathcal{G}_\tau(g_+,g_-) &= g_+ e^{i\tau\hat H_a}\hat G_ae^{-i\tau\hat H_a}{\bullet} - {\bullet}e^{i\tau\hat H_a}\hat G_ae^{-i\tau\hat H_a}g_-.
\end{align}
\end{subequations}
Essentially, Eq.~\eqref{eq:appx:Q_open_env} is of the same form as Eq.~\eqref{eq:appx:Q_obs_super-prop_form} from appendix~\ref{appx:quasi-average_form}.

Next, assume (i) the stationary initial state in $B$, $[\hat\rho_b,\hat H_b] = 0$, and (ii) the weak $AB$ coupling regime,
\begin{align}
    \mu\tau_b \ll 1,
\end{align}
where $\tau_b$ is the \textit{correlation time} in $B$---the time scale on which the components of bi-stochastic process $G_\pm(t)$ decorrelate; in more formal terms, if $\tau_b$ is finite, then it is defined as such a length of time that
\begin{align*}
    \begin{aligned}
        \langle\cdots G_+(t+\tau)G_+(t)\cdots\rangle&\\
        \langle\cdots G_+(t+\tau)G_-(t)\cdots\rangle&\\
        \langle\cdots G_-(t+\tau)G_+(t)\cdots\rangle&\\
        \langle\cdots G_-(t+\tau)G_-(t)\cdots\rangle&
    \end{aligned}
    \xrightarrow{\tau \gg \tau_b}
    \begin{aligned}
        \langle\cdots G_+(t+\tau)\rangle\langle G_+(t)\cdots\rangle&\\
        \langle\cdots G_+(t+\tau)\rangle\langle G_-(t)\cdots\rangle&\\
        \langle\cdots G_-(t+\tau)\rangle\langle G_+(t)\cdots\rangle&\\
        \langle\cdots G_-(t+\tau)\rangle\langle G_-(t)\cdots\rangle&
    \end{aligned},
\end{align*}
i.e., when the distance between consecutive time arguments in a moment is larger than $\tau_b$, the moment factorizes.

One of the effects of weak coupling is the decorrelation of propagators $\Omega$ in between the consecutive projectors $\mathcal{P}$,
\begin{align}\label{eq:appx:almost_QRF}
    \operatorname{tr}_a\!\left[\left\langle\prod_{i=n}^1 \mathcal{P}_{t_i}(f_i,f_{-i})\Omega[G_+,G_-](t_i,t_{i-1})\right\rangle\hat\rho_a\right]\! \approx
    \operatorname{tr}_a\!\left[\left(\prod_{i=n}^1 \mathcal{P}_{t_i}(f_i,f_{-i})\left\langle\Omega[G_+,G_-](t_i,t_{i-1})\right\rangle\right)\hat\rho_a\right].
\end{align}
Indeed, if $t_{i}-t_{i-1} \sim \tau_b$, then $\Omega[G_+,G_-](t_i,t_{i-1})\approx \bullet$ (the super-operator identity) because $\mu(t_i-t_{i-1})\ll 1$, therefore, a non-negligible contribution comes only from propagators for which $t_i - t_{i-1} \gg \tau_b$. But this means that there is enough distance between any two non-negligible $\Omega$'s for $(G_+(t),G_-(t))$ to decorrelate, and so, the bi-average factorizes as in Eq.~\eqref{eq:appx:almost_QRF}.

The other effect is the significant simplification of propagators. Each bi-averaged evolution segment in Eq.~\eqref{eq:appx:QRF},
\begin{align}\label{eq:appx:propagator}
    \left\langle\Omega[G_+,G_-](t_{i},t_{i-1})\right\rangle = \Big\langle Te^{-i\mu\int_{t_{i-1}}^{t_i}\mathcal{G}_\tau(G_+(\tau),G_-(\tau))d\tau} \Big\rangle
\end{align}
formally resembles the dynamical map~\eqref{eq:appx:exp_of_Gamma} that was ``shifted'' in time from interval $(t,0)$ to $(t_i,t_{i-1})$. The weak coupling approximation to the dynamical map (also know as the Born-Markov approximation or Davies approximation) is one of the fundamental results of the theory of open systems (see~\cite{Szankowski_SciPostLecNotes23_,Szankowski_SciPostLecNotes23_sec_5-3_ME-phys-model_} for the derivation carried out in the language of bi-probabilities). The formal resemblance to dynamical map allows us to apply the weak coupling/Born-Markov/Davies approximation to the non-negligible bi-averaged propagators~\eqref{eq:appx:propagator}; as a result we obtain a map belonging to a semi-group (for simplicity we are setting here $\langle G_\pm(t)\rangle = 0$),
\begin{align}
    \left\langle \Omega[G_+,G_-](t_i,t_{i-1})\right\rangle \xrightarrow{\mu\tau_b\ll 1} e^{\mu^2(t_i-t_{i-1})\mathcal{L}},
\end{align}
generated by the GKLS form super-operator,
\begin{subequations}
\begin{align}
    \mathcal{L} &= -i\Big[ \sum_\omega\operatorname{Im}\{\gamma_\omega\}\hat G_\omega^\dagger\hat G_\omega,\bullet\,\Big] 
        + \sum_\omega 2\operatorname{Re}\{\gamma_\omega\}\left(
            \hat G_\omega\bullet\hat G_\omega^\dagger - \frac{1}{2}\{\hat G_\omega^\dagger\hat G_\omega, \bullet\,\}
        \right);\\
    \gamma_\omega &= \int_0^\infty e^{-i\omega s}\langle G_+(s)G_+(0)\rangle ds \quad \text{( and thus, $\operatorname{Re}\{\gamma_\omega\} \geqslant 0$ )};\\
    \hat G_\omega &= \sum_{\alpha,\alpha'}\delta(\omega - \epsilon_\alpha + \epsilon_{\alpha'})|\alpha\rangle\langle \alpha|\hat G_a|\alpha'\rangle\langle\alpha'|
        \quad \text{( where $\hat H_a |\alpha\rangle = \epsilon_\alpha |\alpha\rangle$ ),}
\end{align}
\end{subequations}
which guarantees that the approximated map is completely positive. Plugging this form of the propagator back into Eq.~\eqref{eq:appx:almost_QRF} we arrive at the \textit{quantum regression formula} (QRF)~\cite{Lonigro_PRA22,Lonigro_JPA22},
\begin{align}
\nonumber
    Q_n^{F}(\mathbf{f}_n,\mathbf{f}_{-n},\mathbf{t}_n)\xrightarrow{\mu\tau_b\ll 1}&
        \operatorname{tr}_a\left[\left(\prod_{i=n}^1 \mathcal{P}_{t_i}(f_i,f_{-i})e^{\mu^2(t_i-t_{i-1})\mathcal{L}}\right)\hat\rho_a\right]\\
\label{eq:appx:QRF}
        \equiv& \operatorname{tr}_a\left[\left(\prod_{i=n}^1 \mathcal{P}(f_i,f_{-i})\Lambda(t_i-t_{i-1})\right)\hat\rho_a\right],
\end{align}
where $\mathcal{P}(f_+,f_-) = \hat P_a(f_+)\bullet\hat P_a(f_-) = \mathcal{P}_0(f_+,f_-)$ and
\begin{align}
\nonumber
    \Lambda(t-s) &= e^{(t-s)(-i[\hat H_a,\bullet]+\mu^2\mathcal{L})}\\
    &= \big(e^{-it\hat H_a}{\bullet}e^{it\hat H_a}\big)e^{\mu^2 (t-s)\mathcal{L}}\big(e^{is\hat H_a}{\bullet}e^{-is\hat H_a}\big)
        = \big(e^{-i(t-s)\hat H_a}{\bullet}e^{i(t-s)\hat H_a}\big)e^{\mu^2 (t-s)\mathcal{L}}.
\end{align}

The QRF form~\eqref{eq:appx:QRF} does not satisfy the surrogate field (SF) condition~\eqref{eq:SFR_condition} by default, but it does open new options for suppressing the off-diagonal part of the bi-probability. One such option is to set the map, or rather, its generator $-i[\hat H_a,\bullet]+\mu^2\mathcal{L}$, to be a ``lower block triangular'', i.e.,
\begin{align}\label{eq:appx:lower}
    \forall(f,f_+\neq f_-):\  \mathcal{P}(f,f)\big({-i}[\hat H_a,\bullet]+\mu^2\mathcal{L}\big)\mathcal{P}(f_+,f_-) = 0,
\end{align}
which, of course, implies $\mathcal{P}(f,f)\Lambda(t)\mathcal{P}(f_+,f_-) = 0$ for all $t\geqslant 0$. Now, since each off-diagonal $Q_n^F$ has to involve at least one case of such a super-operator composition (note that $f_{-n} = f_n$ is always true), we conclude that off-diagonal parts of bi-probabilities all vanish individually.

Another option is to have the ``upper block triangular'' generator together with the block diagonal initial density matrix, i.e.,
\begin{align}\label{eq:appx:upper}
    \forall(f,f_+\neq f_-):\  \mathcal{P}(f_+,f_-)\big({-i}[\hat H_a,\bullet]+\mu^2\mathcal{L}\big)\mathcal{P}(f,f) = 0\quad \text{and}\quad  \sum_f\mathcal{P}(f,f)\hat\rho_a =\hat\rho_a  \,\, .
\end{align}
Such a generator (and thus, the map) on its own would not guarantee that all the off-diagonal parts are automatically zero. If there were no constraints placed on the density matrix, it would be possible to pass from non-block diagonal $\hat\rho_a$ to $\mathcal{P}(f_1,f_{-1})$ with impunity. Hence, the assumption $\sum_f\mathcal{P}(f,f)\hat\rho_a = \hat\rho_a$ is needed to close this loophole.

A simple example of a system that satisfies~\eqref{eq:appx:lower} can be found for a two-level $A$ with
\begin{align}
    \hat F = \frac{1}{2}\hat\sigma_z\otimes\hat 1 = \sum_{f = \pm 1/2}f |f\rangle\langle f|\otimes \hat 1;\quad \Lambda(t) = e^{-\frac{1}{2}\gamma t [\hat\sigma_x,[\hat\sigma_x,\bullet\,]]},
\end{align}
resulting in a surrogate field $F(t)$ in the form of random telegraph noise~\cite{Szankowski_SciRep20,Szankowski_SciPostLecNotes23_}---a stochastic process that switches randomly between two values ($\pm 1/2$ in this case) with the rate $\gamma$.

Previous papers investigating Kolmogorovian consistency in the context of sequential measurements~\cite{Smirne_QST19, Strasberg_PRA19,Milz_PRX20} either assume QRF is in effect, or consider the case when $P_n$ can be parameterized with QRF without establishing the relation between actual dynamics and maps appearing in the formula. Therefore, us having arrived at QRF form~\eqref{eq:appx:QRF} as a result of weak coupling approximation gives us an opportunity to compare our results with the conclusions drawn in those previous works. In particular, QRF allows us to reformulate the consistent measurements (CM) condition~\eqref{eq:KCC_condition} as a condition placed on the map (this mimics the approach of~\cite{Smirne_QST19,Milz_PRX20}),
\begin{align}
\nonumber
    &\forall(n\geqslant 1)\forall(0<t_1<\cdots<t_n)\forall(1\leqslant i \leqslant n)\forall(f_1,\ldots,f_{i-1},f_{i+1},\ldots,f_n):\\
\nonumber
    &\left(\prod_{j\neq i}\delta_{f_j,f_{-j}}\right)\sum_{f_i\neq f_{-i}}Q^F_n(\mathbf{f}_n,\mathbf{f}_{-n},\mathbf{t}_n)\\
\nonumber
    &\phantom{=}=
        \operatorname{tr}_a\left[
            \cdots\mathcal{P}(f_{i+1},f_{i+1})
            \Lambda(t_{i+1}-t_i)\left(\sum_{f_i\neq f_{-i}}\mathcal{P}(f_i,f_{-i})\right)\Lambda(t_i-t_{i-1})
            \mathcal{P}(f_{i-1},f_{i-1})\cdots
        \right]\\
\nonumber
    &\phantom{=}=
        \operatorname{tr}_a\left[
            \cdots\mathcal{P}(f_{i+1},f_{i+1})
            \Lambda(t_{i+1}-t_i)\left({\bullet}-\Delta\right)\Lambda(t_i-t_{i-1})
            \mathcal{P}(f_{i-1},f_{i-1})\cdots
        \right]\\
    &\phantom{=}=0,
\end{align}
where $\Delta = \sum_f \mathcal{P}(f,f)$. If the density matrix is block diagonal, $\Delta\hat\rho_a = \hat\rho_a$, then the condition is met when the map satisfies
\begin{align}\label{eq:appx:CM_maps}
    \forall(f,f')\forall(t > t' > 0)&:
    \mathcal{P}(f,f)\Lambda(t)\mathcal{P}(f',f') = \mathcal{P}(f,f)\Lambda(t-t')\Delta\Lambda(t')\mathcal{P}(f',f').
\end{align}
In comparison, it is shown in~\cite{Smirne_QST19,Milz_PRX20} that the Born distributions parameterized with QRF are consistent when (i) the density matrix is block-diagonal, and (ii) the map satisfies the \textit{non coherence generating and detecting} (NCGD) condition,
\begin{align}
    \forall(t>t'>0)&: \Delta\Lambda(t)\Delta = \Delta\Lambda(t-t')\Delta\Lambda(t')\Delta.
\end{align}
Interestingly, the NCGD and CM conditions are, in this case, equivalent. Indeed, on the one hand, when~\eqref{eq:appx:CM_maps} is satisfied it clearly implies NCGD. On the other hand, since $\mathcal{P}(f,f) = \Delta\mathcal{P}(f,f) = \mathcal{P}(f,f)\Delta$, assuming NCGD gives us
\begin{align}
\nonumber
    \mathcal{P}(f,f)\Lambda(t)\mathcal{P}(f',f') &= \mathcal{P}(f,f)\Delta\Lambda(t)\Delta\mathcal{P}(f',f')
        = \mathcal{P}(f,f)\Delta\Lambda(t-t')\Delta\Lambda(t')\Delta\mathcal{P}(f',f')\\
    &= \mathcal{P}(f,f)\Lambda(t-t')\Delta\Lambda(t')\mathcal{P}(f',f').
\end{align}

Finally, we find that the two types of surrogate field-satisfying maps we have listed above can be identified as examples of NCGD map subtypes. According to the classification introduced in~\cite{Smirne_QST19}, the 
lower block triangular map~\eqref{eq:appx:lower} belongs with the \textit{coherence non-activating} maps that satisfy $\Delta \Lambda(t)\Delta = \Delta \Lambda(t)$, and the upper block triangular map~\eqref{eq:appx:upper} is one of the \textit{coherence non-generating} maps that satisfy $\Delta \Lambda(t)\Delta = \Lambda(t)\Delta$.

\subsection{Macro-scale action}

Consider a system living in a Hilbert space with a continuous basis, $\{|x\rangle\}_{x=-\infty}^\infty$, so that the projectors onto eigenspaces of the observable $\hat F$ can be written as
\begin{align}
    \hat P(f) &= \int_{\Gamma(f)}|x\rangle\langle x|dx,
\end{align}
where $\Gamma(f)$ is an interval of a real line corresponding to the eigenvalue $f$. Then, we can rewrite the bi-probability associated with $\hat F$ in the terms of \textit{probability amplitudes} $\phi_t(x|x') = \langle x|{\exp}(-it\hat H)|x'\rangle$,
\begin{align}
\nonumber
    Q_n^F(\mathbf{f}_n,\mathbf{f}_{-n},\mathbf{t}_n) &=
    \left(\prod_{i=1}^n\int_{\Gamma(f_i)}dx_i\int_{\Gamma(f_{-i})}dx_{-i}\right)\int_{-\infty}^\infty dx_0dx_{-0}\,\delta(x_n-x_{-n}) \\ 
\label{eq:appx:Q_amps}
    &\phantom{=}\times
        \langle x_0|\hat\rho|x_{-0}\rangle\left(\prod_{i=0}^{n-1}\phi_{t_{i+1}-t_{i}}(x_{i+1}|x_i)\,\phi_{t_{i+1}-t_{i}}(x_{-(i+1)}|x_{-i})^*\right),
\end{align}
where $t_0 = 0$.

We now assume that the system can be assigned with a macroscopic-scale \textit{action} $S[q] \gg \hbar$, so that the amplitudes can be reformulated as Feynman path integrals~\cite{Brown_2005} with the stationary phase approximation in effect,
\begin{align}\label{eq:appx:path_int}
    \phi_{t-t'}(x|x') &= \int_{q(t') = x'}^{q(t)=x} e^{\frac{i}{\hbar}S[q]}[Dq] \propto e^{\frac{i}{\hbar}S_\mathrm{cls}(x,t|x',t')}.
\end{align}
Here, $S_\mathrm{cls}(x,t|x',t') = S[q_\mathrm{cls}]$ is the action minimized by the path $q_\mathrm{cls}(t)$ that solves the classical equations of motion (with boundary conditions $q_\mathrm{cls}(t) = x$ and $q_\mathrm{cls}(t') = x'$) corresponding to the \textit{least action principle} of the classical mechanics. In other words, $q_\mathrm{cls}(t)$ is the extremum of $S[q]$, and thus, in its vicinity the variation of the phase factor in Eq.~\eqref{eq:appx:path_int} slows down considerably allowing for a brief window of constructive interference; this is the basic explanation why the stationary phase approximation could have been employed here. In the bi-probability~\eqref{eq:appx:Q_amps}, aside the individual probability amplitudes themselves, the interference effects due to macro-scale values of the action also kick in at the point of integration over each interval $\Gamma(f_{\pm i})$, e.g.,
\begin{align}
    \int_{\Gamma(f_i)}e^{\frac{i}{\hbar}S_\mathrm{cls}(x_{i+1},t_{i+1}|x_i,t_i)+\frac{i}{\hbar}S_\mathrm{cls}(x_i,t_i|x_{i-1},t_{i-1})}dx_i.
\end{align}
As a result, such an integral vanishes due to destructive interference unless the interval contains a stationary point of the phase, $x_i^\mathrm{st}\in\Gamma(f_i)$. Of course, the point counts as stationary when
\begin{align}
    0 &= \frac{\partial S_\mathrm{cls}(x_{i+1},t_{i+1}|x_i,t_i)}{\partial x_i}\Big|_{x_i=x_i^\mathrm{st}} + \frac{\partial S_\mathrm{cls}(x_i,t_i|x_{i-1},t_{i-1})\big]}{\partial x_i}\Big|_{x_i=x_i^\mathrm{st}}
    = -p_{i\to i+1}^{t_i} + p_{i-1\to i}^{t_i},
\end{align}
where $p_{i\to i+1}^{t_i}$ ($p_{i-1\to i}^{t_i}$) is the initial (terminal) momentum of the path $i\to i+1$ ($i-1\to i$) that starts at $x_i,t_i$ ($x_{i-1},t_{i-1}$) and ends at $x_{i+1},t_{i+1}$ ($x_{i},t_{i}$). Therefore, the phase is stationary only if there is no jump in the momentum when switching from one path segment to the next. However, even though the segments end/start at the same point, they are otherwise independent, and thus, there is no reason why there should be no momentum discontinuity for an arbitrary value of $x_i$. The one case when the momentum would be continuous at every point is when $x_i$ happens to be a part of the classical path $i-1\to i+1$, i.e., there is no jump when the two paths $i-1\to i$ and $i \to i+1$ are actually parts of a single path going directly from $x_{i-1},t_{i-1}$ to $x_{i+1},t_{i+1}$. Applying the same argument to each integral we conclude that the bi-probability survives the destructive interference only when the arguments $\mathbf{f}_n$ are chosen in such a way that the corresponding intervals $\Gamma(f_n),\ldots,\Gamma(f_1)$ intersect the path $q_\mathrm{cls}(t)$ that solves the classical equations of motion with the boundary conditions $q_\mathrm{cls}(0) = x_0$ and $q_\mathrm{cls}(t_n) = x_n$. Of course, the same goes for $\mathbf{f}_{-n}$ and the classical path with the boundary conditions $q_\mathrm{cls}(0) = x_{-0}$ and $q_\mathrm{cls}(t_n) = x_{-n}$. If the initial density matrix is diagonal, $\langle x_0|\hat\rho|x_{-0}\rangle = \delta(x_0-x_{-0})\rho(x_0)$, then, it follows, that $Q_n^F$ is non-zero only when $\mathbf{f}_{-n} = \mathbf{f}_n$ because $x_{-n} = x_n$ by default and when also $x_{-0}= x_0$ the classical paths passing through $\Gamma(f_i)$'s and $\Gamma(f_{-i})$'s are overlapping. As a result, $Q_n^F(\mathbf{f}_n,\mathbf{f}_{-n},\mathbf{t}_n) = \delta_{\mathbf{f}_n,\mathbf{f}_{-n}}P_n(\mathbf{f}_n,\mathbf{t}_n)$ and the surrogate field (SF) condition~\eqref{eq:SFR_condition} is satisfied.

\bibliographystyle{quantum}
\bibliography{bibliography/CQSP_vs_SFR}

\end{document}